\documentstyle[seceq,epsf,subfigure,wrapfig]{ptptex}

\makeatletter
\@ifundefined{lesssim}{\def\lesssim{\mathrel{\mathpalette\vereq<}}}{}
\@ifundefined{gtrsim}{}{}
\def\vereq#1#2{\lower3pt\vbox{\baselineskip1.5pt \lineskip1.5pt
\ialign{$\m@th#1\hfill##\hfil$\crcr#2\crcr\sim\crcr}}}
\makeatother

\preprintnumber[3cm]{
WU-AP/150/02}

\title{
A Classification of Spherically Symmetric Kinematic Self-Similar Perfect-Fluid Solutions
}

\author{
Hideki~{\sc Maeda},$^{1,}$
\footnote{E-mail: hideki@gravity.phys.waseda.ac.jp}
Tomohiro~{\sc Harada},$^{2,}$
\footnote{E-mail: harada@gravity.phys.waseda.ac.jp}
Hideo~{\sc Iguchi}$^{3,}$\footnote{E-mail: iguchi@th.phys.titech.ac.jp}
\\ and \\
Naoya~{\sc Okuyama}$^{4,}$\footnote{E-mail: okuyama@gravity.phys.waseda.ac.jp}
}

\inst{
$^{1,2,4}$Department of Physics, Waseda University, Tokyo 169-8555, Japan
\\
$^{3}$Department of Physics, Tokyo Institute of Technology, Tokyo 152-8550, Japan
}

\recdate{
}

\abst{
We classify all spherically symmetric spacetimes admitting a kinematic self-similar vector of the second, zeroth or infinite kind. We assume that the perfect fluid obeys either a polytropic equation of state or an equation of state of the form $p=K\mu$, where $p$ and $\mu$ are the pressure and the energy density, respectively, and $K$ is a constant. We study the cases in which the kinematic self-similar vector is not only ``tilted'' but also parallel or orthogonal to the fluid flow. We find that, in contrast to Newtonian gravity, the polytropic perfect-fluid solutions compatible with the kinematic self-similarity are the Friedmann-Robertson-Walker solution and general static solutions. We find three new exact solutions which we call the dynamical solutions (A) and (B) and $\Lambda$-cylinder solution, respectively.
}

\begin{document}

\maketitle

\section{Introduction}
\label{sec:intro}

There is no characteristic scale in Newtonian gravity or general relativity. A set of field equations is invariant under a scale transformation if we assume appropriate matter fields. This implies the existence of scale-invariant solutions to the field equations. Such solutions are called self-similar solutions. Among them, the spherically symmetric self-similar system has been widely researched in the context of both Newtonian gravity and general relativity. Although self-similar solutions are only special solutions of the field equations, it has often been supposed that they play an important role in situations where gravity is an essential ingredient in a spherically symmetric system. In particular, a {\it self-similarity hypothesis} has been proposed, which states that solutions in a variety of astrophysical and cosmological situations may naturally evolve to a self-similar form even if they are initially more complicated~\cite{carr1999}.

Self-similar solutions in Newtonian gravity have been studied in an effort to obtain realistic solutions of gravitational collapse leading to star formation~\cite{penston1969,larson1969,shu1977,hunter1977}. For an isothermal gas cloud, Larson and Penston independently found a self-similar solution, which is called the Larson-Penston solution, describing a gravitationally collapsing sphere~\cite{penston1969,larson1969}. Thereafter, Hunter found a new series of self-similar solutions, and noted that a set of such solutions is infinite and discrete~\cite{hunter1977}. Recent numerical simulations and mode analyses showed that the Larson-Penston solution gives the best description for the central part of a generic collapsing gas sphere~\cite{ti1999,hn1997,hm2000a,hm2000b,fc1993,mh2001}. For a polytropic equation of state, Yahil found the polytropic counterpart of the Larson-Penston solution describing a gravitationally collapsing sphere~\cite{yahil1983} (see also Refs.~\citen{hm2000a,ss1988}). We hereafter refer to these solutions as the polytropic Larson-Penston solutions.

In general relativity, self-similarity is defined by the existence of a homothetic Killing vector field~\cite{CT1971}. Such self-similarity is called the first kind (or homothety). Ori and Piran discovered the general relativistic counterpart of the Larson-Penston self-similar solution together with Hunter's family of self-similar solutions for the perfect fluid obeying an equation of state $p=K\mu$ $(0<K \lesssim0.036)$, where $p$ and $\mu$ are the pressure and the energy density, respectively~\cite{op1987,op1988,op1990}. They observed that a naked singularity forms in this solution for $0<K \lesssim 0.0105$. Harada and Maeda found that generic non-self-similar spherical collapse converges to the general relativistic Larson-Penston solution in an approach to a singularity for $0<K \lesssim0.036$~\cite{hm2001,harada2001}. Since a naked singularity forms for $0<K \lesssim0.0105$, this implies the violation of cosmic censorship in the spherically symmetric case (see also Refs.~\citen{harada1998,hin2002}). This represents the strongest known counterexample against cosmic censorship. It also provides strong evidence for the self-similarity hypothesis in general relativistic gravitational collapse. The question then naturally arises, whether collapsing self-similar solutions with a polytropic equation of state exist in general relativity. If such solutions do exist, they may play an important role in the final stage of generic collapse, as in the $p=K\mu$ case. 

In Newtonian gravity, self-similarity for the polytropic case has a different form of the dimensionless variable from that in the isothermal case, since sound speed is not constant in the former case. The self-similarity coordinate is given by $t^{2-\gamma}/r$ for the polytropic case, and by $t/r$ for the isothermal case. The scaling functions of physical quantities for the former case are also different from those for the latter case~\cite{ss1988}. In general relativity, there exists a natural generalization of homothety called {\it kinematic self-similarity}, which is defined by the existence of a kinematic self-similar vector field~\cite{ch} (see also the earlier related works by Tomita~\cite{tomita}). Kinematic self-similarity is characterized by an index and classified into three kinds: the second, zeroth and infinite kinds.

One can show that an equation of state of the form $p=K\mu$ is the only barotropic one compatible with self-similarity of the first kind~\cite{CT1971}. Self-similar perfect-fluid solutions of the first kind with an equation of state of this form have been classified for the dust case ($K=0$) by Carr~\cite{carr2000} and for the case $0<K<1$ by Carr and Coley~\cite{cc2000} (see also Ref.~\citen{gnu1998}). Special cases in which a homothetic Killing vector is not ``tilted'' (i.e., either parallel or orthogonal to the fluid flow) have also been studied~\cite{coley1991,mcintosh1975}. Kinematic self-similar perfect-fluid solutions have been explored by several authors~\cite{coley1997,bc1998,sbc2001,blvw2002}. Benoit and Coley have studied spherically symmetric spacetimes which admit a kinematic self-similar vector of the second and zeroth kinds~\cite{bc1998}. Sintes, Benoit and Coley have considered spacetimes which admit a kinematic self-similar vector of the infinite kind~\cite{sbc2001}. In these works, the equation of state has not been specified. We have previously investigated spherically symmetric spacetimes which contain a perfect fluid obeying a relativistic polytropic equation of state and admit a kinematic self-similar vector of the second kind in which the kinematic self-similar vector is tilted~\cite{mhio2002}. There, we assumed two kinds of polytropic equation of state in general relativity and showed that such spacetimes must be vacuum in both cases. Although a spherically symmetric spacetime which contains a relativistic polytropic perfect fluid is incompatible with kinematic self-similarity of the second kind, it could be compatible with other kinds of kinematic self-similarities (i.e., the zeroth or infinite kind), or with the case in which a kinematic self-similar vector is parallel or orthogonal to the fluid flow.

In this paper, we extend our previous work in several important ways. We study spacetimes which contain a perfect fluid obeying either a polytropic equation of state or an equation of state $p=K\mu$, and which admit a kinematic self-similar vector field of the second, zeroth or infinite kind. We assume two kinds of relativistic polytropic equations of state and study the case in which a kinematic self-similar vector is not only tilted but also parallel or orthogonal to the fluid flow. 

The organization of this paper is the following. In \S \ref{sec:sssandkss} basic equations in a spherically symmetric spacetime are presented and kinematic self-similarity are briefly reviewed. We treat the cases in which a kinematic self-similar vector is tilted, parallel and orthogonal to the fluid flow in \S \ref{sec:tilted}, \S \ref{sec:parallel} and \S \ref{sec:orthogonal}, respectively. \S \ref{sec:summary} is devoted to summary and discussions. Possible equations of state which are compatible with self-similarity are discussed in appendix A. We adopt units such that $c=1$.

\section{Spherically Symmetric Spacetime and Kinematic Self-Similarity}
\label{sec:sssandkss}

The line element in a spherically symmetric spacetime is given by 

\begin{eqnarray}
ds^2 &=& -e^{2\Phi(t,r)}dt^2+e^{2\Psi(t,r)}dr^2+R(t,r)^2 d\Omega^2,
\end{eqnarray}
where $d\Omega^2=d\theta^2+\sin^2 \theta d\varphi^2$. We consider a perfect fluid as a matter field

\begin{eqnarray}
T_{\mu\nu} &=& p(t,r)g_{\mu\nu}+[\mu(t,r)+p(t,r)]U_{\mu} U_{\nu}, \\
U_{\mu} &=& (-e^{\Phi},0,0,0),
\end{eqnarray}
where $U^{\mu}$ is the four-velocity of the fluid element. We have adopted the comoving coordinates. Then the Einstein equations and the equations of motion for the perfect fluid are reduced to the following simple form:

\begin{eqnarray}
\Phi_r &=& -\frac{p_r}{\mu+p}, \label{basic1}\\
\Psi_t &=& -\frac{\mu_t}{\mu+p}-\frac{2R_t}{R}, \label{basic2}\\
m_r &=& 4\pi \mu R_r R^2,   \label{basic3}\\
m_t &=& -4\pi p R_t R^2, \label{basic4}\\ 
0&=&-R_{tr}+\Phi_r R_{t}+\Psi_t R_r,\label{basic5}\\
m &=& \frac{1}{2G} R(1+e^{-2\Phi}{R_t}^2-e^{-2\Psi}
{R_r}^2 ),\label{basic6}
\end{eqnarray}
where subscripts $t$ and $r$ mean the derivative with respect to $t$ and $r$, respectively and $m(t,r)$ is called the Misner-Sharp mass. We also write the auxiliary equations:
\begin{eqnarray}
&&-\frac{e^{2\Phi}}{R^2}-\left[\left(\frac{R_t}{R}\right)^{2}+
2\frac{R_t}{R} \Psi_t \right] +e^{2\Phi-2\Psi}\left[2\frac{R_{rr}}{R}-
2\frac{R_r}{R} \Psi_r+\left(\frac{R_r}{R}\right)^{2}\right]\nonumber \\
&=&-8\pi G\mu e^{2\Phi},\label{00}\\
&&\frac{e^{2\Psi}}{R^2}+e^{2\Psi-2\Phi}\left[2\frac{R_{tt}}{R}-
2\frac{R_t}{R} \Phi_t+\left(\frac{R_t}{R}\right)^{2}\right] -\left[\left(\frac{R_r}{R}\right)^{2}+2\frac{R_r}{R} \Phi_r\right]\nonumber \\
&=& -8\pi G p e^{2\Psi},\label{11}\\
&&e^{-2\Phi}\left({\Psi}_{tt}+{\Psi}_{t}^2
-\Phi_t \Psi_t+\frac{R_{tt}}{R}+ 
\frac{R_t \Psi_t}{R}-\frac{R_t \Phi_t}{R}\right) \nonumber \\ 
&&-e^{-2\Psi}\left(\Phi_{rr}+\Phi_r^2
-\Phi_r \Psi_r+\frac{R_{rr}}{R}
+\frac{R_r \Phi_r}{R}-\frac{R_r \Psi_r}{R}\right) \nonumber \\
&=&-8\pi G p. \label{22}
\end{eqnarray}
Five of the above nine equations are independent. 

In this paper, we assume the following two kinds of polytropic equations of state. One is 

\begin{eqnarray}
p=K \mu^{\gamma}, \label{eos1}
\end{eqnarray}
where $K$ and $\gamma$ are constants, and the other is~\cite{st}

\begin{eqnarray}
\left\{
\begin{array}{ll}
\displaystyle{p=K n^{\gamma}},\\
\displaystyle{\mu=m_b n+\frac{p}{\gamma-1}},\label{eos2}
\end{array}
\right.
\end{eqnarray}
where the constant $m_b$ and $n(t,r)$ correspond to the mean baryon mass and the baryon number density, respectively. We call the former equation of state (I) (EOS (I)) and the latter equation of state (II) (EOS (II)). Here we assume that $K \ne 0$ and $\gamma \ne 0,1$. We also treat an equation of state

\begin{eqnarray}
p=K \mu. \label{eos3}
\end{eqnarray}
We call this one equation of state (III) (EOS (III)) and assume that $-1\le K \le 1$ and $K \ne 0$. 

We note the properties of EOS (I) and (II). For $\gamma<0$, the fluid suffers from thermodynamical instability. For $0<\gamma<1$, both EOS (I) and (II) are approximated by a dust fluid in high-density regime, since $p/\mu=K\mu^{\gamma-1} \to 0$ for $\mu \to \infty$ for EOS (I) and 

\begin{eqnarray}
\frac{p}{\mu}&=&\frac{Kn^{\gamma-1}}{m_b+\frac{K n^{\gamma-1}}{\gamma-1}} \to 0 \quad \mbox{for $n \to \infty$},
\end{eqnarray}
for EOS (II). For $1<\gamma$, EOS (II) is approximated by EOS (III) with $K=\gamma-1$ in high-density regime since 

\begin{eqnarray}
\mu=m_b n+\frac{K n^{\gamma}}{\gamma-1} \to \frac{K n^{\gamma}}{\gamma-1}=\frac{p}{\gamma-1} \quad \mbox{for $n \to \infty$}.
\end{eqnarray}
In the case of $2<\gamma$ for EOS (II) and $1<\gamma$ for EOS (I), the dominant energy condition can be violated in high-density regime, which will be unphysical. 

In this paper, we introduce a vector field which is called a kinematic self-similar vector field and treat three cases in which it is parallel, orthogonal to the fluid flow and tilted, i.e., neither of them. In a spherically symmetric spacetime, a vector field ${\bf \eta}$ is written in general as 

\begin{eqnarray}
\eta^{\mu}\frac{\partial}{\partial x^{\mu}}=h_1(t,r) \frac{\partial}{\partial t}+h_2(t,r) \frac{\partial}{\partial r},
\end{eqnarray}
where $h_1(t,r)$ and $h_2(t,r)$ are arbitrary functions. $h_2=0$ when ${\bf \eta}$ is parallel to the fluid flow, while $h_1=0$, when ${\bf \eta}$ is orthogonal to the fluid flow. When ${\bf \eta}$ is tilted, both $h_1$ and $h_2$ are non-zero.

A kinematic self-similarity vector ${\bf\xi}$ satisfies the condition

\begin{eqnarray}
{\cal{L}}_{\bf\xi} h_{\mu\nu} &=&2\delta h_{\mu\nu},\label{kss}\\
{\cal{L}}_{\bf\xi} U_{\mu} &=&\alpha U_{\mu},\label{gss}
\end{eqnarray}
where $h_{\mu\nu} =g_{\mu\nu}+U_{\mu}U_{\nu}$ is the projection tensor, ${\cal{L}}_{\bf\xi}$ denotes the Lie differentiation along ${\bf\xi}$ and $\alpha$ and $\delta$ are constants~\cite{ch,coley1997}. The similarity transformation is characterized by the scale-independent ratio, $\alpha/\delta$, which is referred to as the similarity index. 

In the case of $\delta \ne 0$, $\delta$ can be set to unity and the kinematic self-similar vector ${\bf\xi}$ can be written as 

\begin{eqnarray}
\xi^{\mu}\frac{\partial}{\partial x^{\mu}}=(\alpha t+\beta) \frac{\partial}{\partial t}+r \frac{\partial}{\partial r},
\end{eqnarray}
if it is tilted. In the case of $\alpha=1$, which is corresponding to self-similarity of the first kind ($\beta$ can be set to zero), it follows that ${\bf\xi}$ is a homothetic vector and the self-similar variable $\xi$ is given by $r/t$ . In the case of $\alpha=0$, which is corresponding to self-similarity of the zeroth kind ($\beta$ can be rescaled to unity), the self-similar variable is given by 

\begin{eqnarray}
\xi=r e^{-t}.
\end{eqnarray}
In the case of $\alpha \ne 0, 1$, which is corresponding to self-similarity of the second kind ($\beta$ can be set to zero), the self-similar variable is given by 

\begin{eqnarray}
\xi=\frac{r}{(\alpha t)^{1/\alpha}}.
\end{eqnarray}
In the case of $\delta \ne 0$, self-similarity implies that the metric functions can be written as 

\begin{eqnarray}
R=r S(\xi), \quad \Phi=\Phi(\xi), \quad \Psi=\Psi(\xi).\label{finitessform}
\end{eqnarray}
In the special case of $\delta=0$ and $\alpha \ne 0$, the self-similarity is referred to as the infinite kind ($\alpha=1$ is possible). The kinematic self-similar vector ${\bf\xi}$ can be written as 

\begin{eqnarray}
\xi^{\mu}\frac{\partial}{\partial x^{\mu}}=t \frac{\partial}{\partial t}+r \frac{\partial}{\partial r}.
\end{eqnarray}
The self-similar variable is given by 

\begin{eqnarray}
\xi=\frac{r}{t}.
\end{eqnarray}
In the case of $\delta = 0$, self-similarity implies that the metric functions can be written as 

\begin{eqnarray}
R=S(\xi), \quad e^{\Phi}=e^{\Phi(\xi)}, \quad e^{\Psi}=e^{\Psi(\xi)}/r.
\end{eqnarray}
It is noted that ${\bf \xi}$ is a Killing vector in the case of $\delta=0$ and $\alpha = 0$. If the kinematic self-similar vector ${\bf \xi}$ is parallel to the fluid flow, it is written as

\begin{eqnarray}
\xi^{\mu}\frac{\partial}{\partial x^{\mu}}=f(t) \frac{\partial}{\partial t},\label{xiparallel}
\end{eqnarray}
where $f(t)$ is an arbitrary function and the self-similar variable is then $r$. If the kinematic self-similar vector ${\bf \xi}$ is orthogonal to the fluid flow, it is written as

\begin{eqnarray}
\xi^{\mu}\frac{\partial}{\partial x^{\mu}}=g(r) \frac{\partial}{\partial r},\label{xiorthogonal}
\end{eqnarray}
where $g(r)$ is an arbitrary function and the self-similar variable is then $t$. 

Mathematical and physical properties of spacetimes which admit a kinematic self-similar vector of the ``finite'' kind, i.e., the second and zeroth kinds, and contain a perfect fluid have been discussed by Coley~\cite{coley1997}. It is noted that they have assumed ``physical'' self-similarity, i.e.,

\begin{eqnarray}
{\cal{L}}_{\bf\xi} \mu =a\mu,\quad {\cal{L}}_{\bf\xi} p =bp, \label{pss}
\end{eqnarray}
where $a$ and $b$ are constants, in addition to kinematic self-similarity as represented by equations (\ref{kss}) and (\ref{gss}). The spacetimes which admit a kinematic self-similar vector of the infinite kind and contain a perfect fluid have been studied by Sintes, Benoit and Coley~\cite{sbc2001}. Benoit and Coley have investigated spherically symmetric spacetimes which admit a tilted kinematic self-similar vector field of the second or zeroth kind and contain a perfect fluid without specifying the equation of state~\cite{bc1998}.

\section{Tilted case}
\label{sec:tilted}

\subsection{Self-similarity of the the second kind}
In the case of self-similarity of the second kind, the Einstein equations imply that the quantities $m, \mu$ and $p$ must be of the form 

\begin{eqnarray}
2Gm&=&r\left[M_1(\xi)+\frac{r^2}{t^2}M_2(\xi)\right],\label{2ndm}\\
8\pi G \mu&=&\frac{1}{r^2}\left[W_1(\xi)+\frac{r^2}{t^2}W_2(\xi)\right],\label{2ndmu}\\
8\pi G p&=&\frac{1}{r^2}\left[P_1(\xi)+\frac{r^2}{t^2}P_2(\xi)\right],\label{2ndp}
\end{eqnarray}
where $\xi=r/(\alpha t)^{1/\alpha}$. A set of ordinary differential equations is obtained when one demands that the Einstein equations and the equations of motion for the matter field are satisfied for the $O[(r/t)^0]$ and $O[(r/t)^2]$ terms separately. The equations for a perfect fluid (\ref{basic1})-(\ref{11}) reduce to the following:

\begin{eqnarray}
M_1+M_1'&=&W_1S^2(S+S'),\label{2nd1}\\
3M_2+M_2'&=&W_2S^2(S+S'),\label{2nd2}\\
M_1'&=&-P_1S^2S',\label{2nd3}\\
2\alpha M_2+M_2'&=&-P_2S^2S',\label{2nd4}\\
M_1&=&S[1-e^{-2\Psi}(S+S')^2],\label{2nd5}\\
\alpha^2 M_2&=&SS'^2 e^{-2\Phi},\label{2nd6}\\
(P_1+W_1)\Phi'&=&2P_1-P_1',\label{2nd7}\\
(P_2+W_2)\Phi'&=&-P_2',\label{2nd8}\\
W_1'S&=&-(P_1+W_1)(\Psi'S+2S'),\label{2nd9}\\
(2\alpha W_2+W_2')S&=&-(P_2+W_2)(\Psi'S+2S'),\label{2nd10}\\
S''+S'&=&S'\Phi'+(S+S')\Psi',\label{2nd11}\\
S'(S'+2\Psi'S)&=&\alpha^2W_2 S^2 e^{2\Phi}, \label{2nd00a}\\
2S(S''+2S')-2\Psi'S(S+S')&=&-S'^2-S^2+e^{2\Psi}(1-W_1S^2),\label{2nd00b}\\
2S(S''+\alpha S'-\Phi'S')+S'^2&=&-\alpha^2 P_2S^2 e^{2\Phi},\label{2nd11a}\\
(S+S')(S+S'+2\Phi'S)&=&(1+P_1S^2)e^{2\Psi},\label{2nd11b}
\end{eqnarray}
where the prime denotes the derivative with respect to $\ln \xi$. 

In a vacuum case, while the Minkowski spacetime is compatible with all $\alpha$, the Schwarzschild spacetime is compatible only with $\alpha=3/2$, since equations (\ref{2nd2}) and (\ref{2nd4}) are degenerated so that $M_2 \ne 0$ is possible only in this case. The Schwarzschild spacetime in the Lemaitre's choice of coordinates is written as

\begin{equation}
ds^2=-dt^2+r_g^{2/3}\left(\frac{d\rho^2}{\left[\frac32(\rho-t)\right]^{2/3}}+\left[\frac32(\rho-t)\right]^{4/3}d\Omega^2\right),
\end{equation}
where $r_g$ is a constant and the Schwarzschild radius corresponds to $r_g=(3/2)(\rho-t)$~\cite{LL}. Changing the radial coordinate as $\rho=r^{3/2}$, the metric in the form of (\ref{finitessform}) for $\alpha=3/2$ can be obtained:

\begin{equation}
ds^2=-dt^2+r_g^{2/3}\left(\frac{(9/4)dr^2}{\left[\frac32(1-t/r^{3/2})\right]^{2/3}}+r^2\left[\frac32(1-t/r^{3/2})\right]^{4/3}d\Omega^2\right).
\end{equation}

\subsubsection{EOS (I) and (II)}
Subtracting equation (\ref{2nd11b}) from equation (\ref{2nd00b}) and eliminating $S''$ with using equation (\ref{2nd11}), we obtain

\begin{eqnarray}
2\Phi'&=&(P_1+W_1)e^{2\Psi}. \label{2ndkey1}
\end{eqnarray}
Then equations (\ref{2nd7}) and (\ref{2nd8}) result in

\begin{eqnarray}
e^{2\Psi}(P_1+W_1)^2&=&4P_1-2P_1',\label{2ndkey2}\\
e^{2\Psi}(P_1+W_1)(P_2+W_2)&=&-2P_2'.\label{2ndkey3}
\end{eqnarray}

If a perfect fluid obeys EOS (I) for $K \ne 0$ and $\gamma \ne 0, 1$, we find from equations (\ref{2ndmu}) and (\ref{2ndp}) that

\begin{eqnarray}
\alpha=\gamma, \quad P_1=W_2=0,\quad P_2 =\frac{K}{(8\pi G)^{\gamma-1}\gamma^2}{\xi}^{-2\gamma}W_1^{\gamma}, \quad \mbox{[case (A)]} \label{casea}
\end{eqnarray}
or
\begin{eqnarray}
\alpha=\frac{1}{\gamma},\quad P_2=W_1=0,\quad P_1 =\frac{K}{(8\pi G)^{\gamma-1}\gamma^{2\gamma}}{\xi}^2 W_2^{\gamma}. \quad \mbox{[case (B)]} \label{caseb}
\end{eqnarray}
If a perfect fluid obeys EOS (II) for $K \ne 0$ and $\gamma \ne 0, 1$, we find from equations (\ref{2ndmu}) and (\ref{2ndp}) that

\begin{equation}
\alpha=\gamma,\quad P_1=0,\quad P_2 =\frac{K}{m_b^{\gamma}(8\pi G)^{\gamma-1}\gamma^2}{\xi}^{-2\gamma}W_1^{\gamma}=(\gamma-1)W_2, \quad \mbox{[case (C)]} \label{casec}
\end{equation}
or
\begin{equation}
\alpha=\frac{1}{\gamma},\quad P_2=0,\quad P_1 =\frac{K}{m_b^{\gamma}(8\pi G)^{\gamma-1}\gamma^{2\gamma}}{\xi}^2 W_2^{\gamma}=(\gamma-1)W_1. \quad \mbox{[case (D)]} \label{cased}
\end{equation}

$P_1=0$ directly implies $W_1=0$ from equation (\ref{2ndkey2}), while $P_2=0$ implies $(P_1+W_1)W_2=0$ from (\ref{2ndkey3}), which results in $W_2=0$ for cases (B) and (D). Therefore the spacetime must be vacuum for all possible cases.

\subsubsection{EOS (III)}
If a perfect fluid obeys EOS (III), we find from equations (\ref{2ndmu}) and (\ref{2ndp}) that 

\begin{eqnarray}
P_1=K W_1, \quad P_2=KW_2. \quad \mbox{[case (E)]} \label{casee} 
\end{eqnarray}
$K \ne -1$ can be concluded, since $K=-1$ implies $P_1=W_1=0$ from equations (\ref{2nd7}) and (\ref{2nd9}) and also $P_2=W_2=0$ from equations (\ref{2nd8}) and (\ref{2nd10}), which means a vacuum spacetime. For $K \ne -1$, $W_1W_2=0$ can be proved. Assuming that $W_1 \ne 0$ and $W_2 \ne 0$, we find from equations (\ref{2nd9}) and (\ref{2nd10}) that $W_1'/W_1=2\alpha+W_2'/W_2$, while equations (\ref{2nd7}) and (\ref{2nd8}) give that $2-W_1'/W_1=-W_2'/W_2$. These two equations contradict the assumption of $\alpha \ne 1$. Hence $W_1W_2=0$ is concluded.

In the case of $W_1=P_1=0$ and $W_2 \ne 0$, $M_1=0$ can be obtained from equations (\ref{2nd1}) and (\ref{2nd3}), and then equation (\ref{2nd5}) gives $(S+S')^2=e^{2\Psi}$. Equation (\ref{2ndkey1}) gives $\Phi'=0$, and then equation (\ref{2nd8}) requires $P_2$ to be constant, which implies that $W_2$ is also constant. We set $e^{\Phi}=c_0$ and $W_2=w_0$, where $c_0$ and $w_0$ are constants. Equations (\ref{2nd2}), (\ref{2nd4}) and (\ref{2nd6}) give the evolution equation for $S$:

\begin{eqnarray}
\frac{3-2\alpha}{\alpha^2 c_0^2}\left(\frac{S'}{S}\right)^2-(1+K)w_0\frac{S'}{S}-w_0=0. \label{evols}
\end{eqnarray}
The solution is $S=s_0 \xi^q$, where $s_0$ and $q$ are constants. $q=-2\alpha/[3(1+K)]$ can be obtained from equation (\ref{2nd10}). The resulting solution is 

\begin{eqnarray}
e^{\Phi}&=&c_0,\\
e^{\Psi}&=&s_0 \left|1-\frac{2\alpha}{3(1+K)}\right|\xi^{-\frac{2\alpha}{3(1+K)}},\label{psi}\\
S&=&s_0\xi^{-\frac{2\alpha}{3(1+K)}},\\
M_1&=&0, \quad M_2=\frac{4s_0^3}{9c_0^2(1+K)^2}\xi^{-\frac{2\alpha}{1+K}},\\
P_1&=&W_1=0,\\
P_2&=&KW_2=\frac{4K}{3c_0^2(1+K)^2}.
\end{eqnarray}
This is the flat Friedmann-Robertson-Walker (FRW) solution. Equation (\ref{psi}) implies $\alpha \ne 3(1+K)/2$. With the coordinate transformations $t'=c_0 t$ and $r'=(c_0/\alpha)^{-2/[3(1+K)]}s_0 r^{1-\frac{2\alpha}{3(1+K)}}$, the solution in a more usual form 

\begin{eqnarray}
ds^2&=&-dt'^2+t'^{\frac{4}{3(1+K)}}(dr'^2+r'^2d\Omega^2),\\
2Gm&=&\frac{4}{9(1+K)^2}\frac{r'^3}{t'^{\frac{2K}{1+K}}},\\
8\pi G p&=&8\pi G K\mu=\frac{4K}{3(1+K)^2t'^2},
\end{eqnarray}
is obtained. 

In the case of $W_2=P_2=0$ and $W_1 \ne 0$, equations (\ref{2nd9}) and (\ref{2nd00a}) give 

\begin{eqnarray}
\frac{S'}{S}\left(3\frac{S'}{S}+\frac{2}{1+K}\frac{W_1'}{W_1}\right)=0.\label{eqs1}
\end{eqnarray}
Eliminating $S''-S'\Phi'$ and $\Psi'$ from equation (\ref{2nd11a}) with using equations (\ref{2nd11}) and (\ref{2nd9}), respectively, we obtain  

\begin{eqnarray}
2\left[(\alpha-1)\frac{S'}{S}-\left(1+\frac{S'}{S}\right)\left(2\frac{S'}{S}+\frac{1}{1+K}\frac{W_1'}{W_1}\right)\right]+\left(\frac{S'}{S}\right)^2=0.\label{secondkey}
\end{eqnarray}
If $3S'/S+[2/(1+K)]W_1'/W_1=0$, equation (\ref{secondkey}) results in $(2\alpha-3)S'=0$ so that $S'=0$ or $\alpha=3/2$.

If $S'=0$, the resulting solution is 

\begin{eqnarray}
e^{\Phi}&=&c_0 \xi^{2K/(1+K)},\label{static1}\\
e^{\Psi}&=&\frac{s_0}{\sqrt{1-w_0 s_0^2}},\\
S&=&s_0,\\
M_1&=&w_0 s_0^3,\quad M_2=0,\\
P_1&=&KW_1=Kw_0,\\
P_2&=&W_2=0,\label{static2}
\end{eqnarray}
where $c_0$, $s_0$ and $w_0$ are constants and $4K=w_0 s_0^2(K^2+6K+1)$. $-1<K<2\sqrt{2}-3$ and $0<K \le 1$ must be satisfied for the positive energy density. With the coordinate transformations $t'=c_0s_0^{-2K/(1+K)}\alpha^{-2K/[\alpha(1+K)]}(1-2K/[\alpha(1+K)])^{-1}t^{1-\frac{2K}{\alpha(1+K)}}$ and $r'=s_0 r$, the solution is found to be a static solution:

\begin{eqnarray}
ds^2&=&-r'^{4K/(1+K)} dt'^2+\frac{K^2+6K+1}{(1+K)^2}dr'^2+r'^2 d\Omega^2,\\
2Gm&=&\frac{4K}{K^2+6K+1}r',\\
8\pi G p&=&8\pi G K\mu=\frac{4K^2}{(K^2+6K+1)r'^2}.
\end{eqnarray}
This solution is singular at the physical center $r'=0$. We call this solution the singular static solution. It is noted that this solution is also of the first kind~\cite{cc2000}.


For $\alpha = 3/2$, equations (\ref{2nd2}) and (\ref{2nd4}) are degenerate. Equation (\ref{2nd7}) and $3S'/S+[2/(1+K)]W_1'/W_1=0$ give

\begin{eqnarray}
\Phi'=\frac{2K}{1+K}+\frac{3K}{2}\frac{S'}{S}.\label{Phider}
\end{eqnarray}
Equations (\ref{Phider}) and (\ref{2nd11a}) give the evolution equation for $S$:

\begin{eqnarray}
2(1+K)y'+(3-K)y+3(1-K)(1+K)y^2=0,\label{evolsdynama}
\end{eqnarray}
where $y \equiv S'/S$. This equation can be integrated as

\begin{eqnarray}
S=s_1|1-s_0 \xi^{-\frac{3-K}{2(1+K)}}|^{\frac{2}{3(1-K)}},\label{dynaa1st}
\end{eqnarray}
where $s_0$ and $s_1$ are integration constants. An identity related to $M_1$ can be obtained from equations (\ref{2nd1}), (\ref{2nd3}) and (\ref{2nd5}):

\begin{eqnarray}
W_1S(S+S')+KW_1SS'=1-e^{-2\Psi}(S+S')^2. \label{2ndpart1}
\end{eqnarray}
Eliminating $\Phi'$ from equation (\ref{2nd11b}) with using equations (\ref{2nd11a}), we obtain

\begin{equation}
2(S+S')(SS''+2SS'+S'^2)=S'(1+KW_1S^2)e^{2\Psi}.\label{2ndpart2}
\end{equation}
From equations (\ref{evolsdynama}), (\ref{2ndpart1}) and (\ref{2ndpart2}), we obtain

\begin{equation}
e^{2\Psi}=\frac{(K^2+6K+1)S^2+(1+K)(9K+2)SS'+(3K+1)(1+K)^2S'^2}{(1+K)^2},\label{dynaa2nd}
\end{equation}
and

\begin{equation}
W_1=\frac{P_1}{K}=\frac{K[4+3(1+K)S'/S]}{(K^2+6K+1)S^2+(1+K)(9K+2)SS'+(3K+1)(1+K)^2S'^2}.
\end{equation}
$\exp(\Phi)$, $M_1$ and $M_2$ can be obtained from equations (\ref{Phider}), (\ref{2nd5}) and (\ref{2nd6}), respectively:

\begin{eqnarray}
e^{\Phi}&=&c_0\xi^{2K/(1+K)}S^{3K/2},\\
M_1&=&S[1-e^{-2\Psi}(S+S')^2],\\
M_2&=&\frac49 e^{-2\Phi}SS'^2,\label{dynaalast}
\end{eqnarray}
where $c_0$ is an integration constant. Since $3S'/S+[2/(1+K)][W_1'/W_1]=0$ gives $S^3 \propto W_1^{-2/(1+K)}$, this solution is singular at the physical center $S=0$. For $s_0>0$, this solution represents collapsing shells, since $S$ decreases as $\xi$ decreases from $\xi=+\infty$ which means that $t$ increases from $t=+0$ for constant $r$. A shell for some constant $r$ collapses to a singularity $S=0$ at a finite time $t_1=(2/3)s_0^{-3(1+K)/(3-K)}r^{3/2}$. For $s_0<0$, this solution represents expanding shells, since $S$ increases as $\xi$ decreases from $\xi=+\infty$ which means that $t$ increases from $t=+0$ for constant $r$. In this case, $S$ cannot be zero. We call this the dynamical solution (A).


\subsection{Self-similarity of the the zeroth kind}
In the case of self-similarity of the zeroth kind, the Einstein equations imply that the quantities $\mu, p$ and $m$ must be of the form 

\begin{eqnarray}
2Gm&=&r[M_1(\xi)+r^2 M_2(\xi)],\label{zerom}\\
8\pi G \mu&=&\frac{1}{r^2}[W_1(\xi)+r^2W_2(\xi)],\label{zeromu}\\
8\pi G p&=&\frac{1}{r^2}[P_1(\xi)+r^2P_2(\xi)],\label{zerop}
\end{eqnarray}
where $\xi=r e^{-t}$. A set of ordinary differential equations is obtained when one demands that the Einstein equations and the equations of motion for the matter field are satisfied for the $O[r^0]$ and $O[r^2]$ terms separately. The resulting equations for a perfect fluid (\ref{basic1})-(\ref{11}) reduce to the following:

\begin{eqnarray}
M_1+M_1'&=&W_1S^2(S+S'),\label{zero1}\\
3M_2+M_2'&=&W_2S^2(S+S'),\label{zero2}\\
M_1'&=&-P_1S^2S',\label{zero3}\\
M_2'&=&-P_2S^2S',\label{zero4}\\
M_1&=&S[1-e^{-2\Psi}(S+S')^2],\label{zero5}\\
M_2&=&SS'^2 e^{-2\Phi},\label{zero6}\\
(P_1+W_1)\Phi'&=&2P_1-P_1',\label{zero7}\\
(P_2+W_2)\Phi'&=&-P_2',\label{zero8}\\
W_1'S&=&-(P_1+W_1)(\Psi'S+2S'),\label{zero9}\\
W_2'S&=&-(P_2+W_2)(\Psi'S+2S'),\label{zero10}\\
S''+S'&=&S'\Phi'+(S+S')\Psi',\label{zero11}\\
S'(S'+2\Psi'S)&=&W_2 S^2 e^{2\Phi}, \label{zero00a}\\
2S(S''+2S')-2\Psi'S(S+S')&=&-S'^2-S^2+e^{2\Psi}(1-W_1S^2),\label{zero00b}\\
2S(S''-\Phi'S')+S'^2&=&-P_2S^2 e^{2\Phi},\label{zero11a}\\
(S+S')(S+S'+2\Phi'S)&=&(1+P_1S^2)e^{2\Psi},\label{zero11b}
\end{eqnarray}
where the prime denotes the derivative with respect to $\ln \xi$. In a vacuum case, the Minkowski spacetime can be obtained since $M_1=M_2=0$.

\subsubsection{EOS (I) and (II)}
Subtracting equation (\ref{zero11b}) from equation (\ref{zero00b}) and eliminating $S''$ with using equation (\ref{zero11}), we obtain

\begin{eqnarray}
2\Phi'=(P_1+W_1)e^{2\Psi}.\label{zerokey1}
\end{eqnarray}
Then equations (\ref{zero7}) and (\ref{zero8}) result in

\begin{eqnarray}
e^{2\Psi}(P_1+W_1)^2&=&4P_1-2P_1',\label{zerokey2}\\
e^{2\Psi}(P_1+W_1)(P_2+W_2)&=&-2P_2'\label{zerokey3}.
\end{eqnarray}

If a perfect fluid obeys EOS (I), we find from equations (\ref{zeromu}) and (\ref{zerop}) that

\begin{eqnarray}
P_1=W_1=0,\quad P_2 =K(8 \pi G)^{1-\gamma}W_2^\gamma. \quad \mbox{[case (A)]}\label{casea2}
\end{eqnarray}
If a perfect fluid obeys EOS (II), we find from equations (\ref{zeromu}) and (\ref{zerop}) that

\begin{eqnarray}
P_1=W_1=0,\quad P_2 =\frac{K}{(8\pi G)^{\gamma-1} m_b^{\gamma}}\left(W_2-\frac{P_2}{\gamma-1}\right)^\gamma. \quad \mbox{[case (B)]}\label{caseb2}
\end{eqnarray}

Since $P_1=W_1=0$, $M_1=0$ can be obtained from equations (\ref{zero1}) and (\ref{zero3}). Equation (\ref{zero5}) gives $(S+S')^2=e^{2\Psi}$. From equations (\ref{zerokey1}) and (\ref{zerokey3}), $P_1=W_1=0$ implies $\Phi'=0$ and $P_2=p_0$, where $p_0$ is a constant, which means that $W_2=w_0$, where $w_0$ is a constant. We set $e^{\Phi}=c_0$, where $c_0$ is a constant. Then equation (\ref{zero10}) gives $(\Psi'S+2S')(P_2+W_2)=0$. $P_2+W_2=0$ means $p+\mu=0$, which will be treated in the next subsection. If $\Psi'S+2S'=0$, the equations (\ref{zero2}), (\ref{zero4}) and (\ref{zero6}) give the evolution equation for $S$;

\begin{eqnarray}
\frac{3}{c_0^2}\left(\frac{S'}{S}\right)^2-(p_0+w_0)\frac{S'}{S}-w_0=0. \label{evols2}
\end{eqnarray}
The solution is $S=s_0 \xi^q$ where $q$ is a constant. $q=0$ can be obtained from equations $(S+S')^2=e^{2\Psi}$ and $\Psi'S+2S'=0$, which means that $S=s_0$ and $P_2=W_2=0$. Therefore the spacetime must be vacuum.

\subsubsection{EOS (III)}
If a perfect fluid obeys EOS (III), we find from equations (\ref{zeromu}) and (\ref{zerop}) that 

\begin{eqnarray}
P_1=K W_1, \quad P_2=KW_2. \quad \mbox{[case (C)]}\label{casec2} 
\end{eqnarray}
$W_1 W_2=0$ can be proved. Assuming that $W_1 \ne 0$ and $W_2 \ne 0$, we find from equations (\ref{zero9}) and (\ref{zero10}) that $W_1'/W_1=W_2'/W_2$, while equations (\ref{zero7}) and (\ref{zero8}) give that $2-W_1'/W_1=-W_2'/W_2$. These two equations contradict each other. Hence $W_1W_2=0$ can be concluded.

In the case of $W_1=P_1=0$ and $W_2 \ne 0$, the discussion in the previous subsection applies and gives that 

\begin{eqnarray}
e^{2\Phi}&=&c_0^2,\\
e^{2\Psi}&=&(S+S')^2,\\
M_1&=&0,\\
M_2&=&c_0^{-2}SS'^2 ,\\
P_2&=&-W_2=p_0,
\end{eqnarray}
where $c_0$ and $p_0$ are constants. The evolution equation for $S$ is

\begin{eqnarray}
\frac{3}{c_0^2}\left(\frac{S'}{S}\right)^2=-p_0,
\end{eqnarray}
with a general solution

\begin{eqnarray}
S=s_0 \xi^{-\sqrt{\frac{c_0^2(-p_0)}{3}}}.
\end{eqnarray}
This spacetime is the de-Sitter spacetime with the metric 

\begin{equation}
ds^2=-c_0^2dt^2+e^{2\sqrt{-c_0^2p_0/3}t}[(1-\sqrt{-c_0^2p_0/3})^2s_0^2 r^{-2\sqrt{-c_0^2p_0/3}}dr^2+s_0^2r^{2(1-\sqrt{-c_0^2p_0/3})}d\Omega^2].
\end{equation}
With the coordinate transformations $t'=c_0t$ and $r'=s_0r^{1-\sqrt{-c_0^2p_0/3}}$, the solution in a more usual form 

\begin{eqnarray}
ds^2&=&-dt'^2+e^{2\sqrt{-p_0/3}t'}(dr'^2+r'^2d\Omega^2),\\
2Gm&=&-\frac{p_0}{3}r'^3e^{3\sqrt{-p_0/3}t'},\\
8\pi G p&=&-8\pi G \mu=8\pi G p_0,
\end{eqnarray}
is obtained.

In the case of $W_2=P_2=0$ and $W_1 \ne 0$, equations (\ref{zero2}) and (\ref{zero4}) give $M_2=0$, and then $S'=0$ can be found from equation (\ref{zero6}). If $K=-1$, $P_1=-W_1=p_0$ is obtained from equation (\ref{zero9}) where $p_0$ is a constant. Equation (\ref{zero7}) then gives $p_0=0$ which means that the spacetime must be vacuum. For $K \ne -1$, the resulting solution is 

\begin{eqnarray}
e^{\Phi}&=&c_0 \xi^{2K/(1+K)},\label{static1zero}\\
e^{\Psi}&=&\frac{s_0}{\sqrt{1-w_0 s_0^2}},\\
S&=&s_0,\\
M_1&=&w_0 s_0^3,\quad M_2=0,\\
P_1&=&KW_1=Kw_0,\\
P_2&=&W_2=0,\label{static2zero}
\end{eqnarray}
where $c_0$, $s_0$ and $w_0$ are constants and $4K=w_0 s_0^2(K^2+6K+1)$. With the coordinate transformations $t'=c_0s_0^{-2K/(1+K)}[-(1+K)/(2K)]e^{-\frac{2K}{1+K}t}$ and $r'=s_0 r$, this solution represents the same singular static solution as equations (\ref{static1})-(\ref{static2}). $-1<K<2\sqrt{2}-3$ and $0<K \le 1$ must be satisfied for the positive energy density. 

\subsection{Self-similarity of the infinite kind}

In the case of self-similarity of the infinite kind, the Einstein equations imply that the quantities $\mu, p$ and $m$ must be of the form 

\begin{eqnarray}
2Gm&=&M_1(\xi)/t^2+M_2(\xi),\label{infm}\\
8\pi G \mu&=&W_1(\xi)/t^2+W_2(\xi),\label{infmu}\\
8\pi G p&=&P_1(\xi)/t^2+P_2(\xi),\label{infp}
\end{eqnarray}
where $\xi=r/t$. A set of ordinary differential equations is obtained when one demands that the Einstein equations and the equations of motion for the matter field are satisfied for the $O[t^0]$ and $O[t^{-2}]$ terms separately. The resulting equations for a perfect fluid (\ref{basic1})-(\ref{22}) reduce to the following:

\begin{eqnarray}
M_1'&=&W_1S^2 S',\label{inf1}\\
M_2'&=&W_2S^2 S'.\label{inf2}\\
2M_1+M_1'&=&-P_1S^2S',\label{inf3}\\
M_2'&=&-P_2S^2S',\label{inf4}\\
M_1&=&e^{-2\Phi}S S'^2,\label{inf5}\\
M_2&=&S(1-e^{-2\Psi}S'^2),\label{inf6}\\
(P_1+W_1)\Phi'&=&-P_1',\label{inf7}\\
(P_2+W_2)\Phi'&=&-P_2',\label{inf8}\\
(2W_1+W_1')S&=&-(P_1+W_1)(\Psi'S+2S'),\label{inf9}\\
W_2'S&=&-(P_2+W_2)(\Psi'S+2S'),\label{inf10}\\
S''&=&S'(\Phi'+\Psi')\label{inf11},\\
W_1S^2e^{2\Phi}&=&S'(2\Psi'S+S'),\label{inf00a}\\
(1-W_2S^2)e^{2\Psi}&=&2SS''+S'^2-2\Psi'S'S,\label{inf00b}\\
(1+P_2S^2)e^{2\Psi}&=&S'(2\Phi'S+S'),\label{inf11a}\\
-P_1S^2e^{2\Phi}&=&2S''S+2SS'(1-\Phi')+S'^2,\label{inf11b}\\
-P_1 Se^{2\Phi}&=&S''+\Psi''S+(1-\Phi'+\Psi')(S'+\Psi'S),\label{inf22a}\\
-P_2 Se^{2\Psi}&=&-S''-\Phi''S+(\Psi'-\Phi')(S'+\Phi'S),\label{inf22b}
\end{eqnarray}
where the prime denotes the derivative with respect to $\ln \xi$. From equations (\ref{inf11}), (\ref{inf00a}) and (\ref{inf11b}), 

\begin{eqnarray}
(P_1+W_1)Se^{2\Phi}=-2S',\label{infkeya}
\end{eqnarray}
can be obtained, while from equations (\ref{inf11}), (\ref{inf00b}) and (\ref{inf11a}), 

\begin{eqnarray}
P_2+W_2=0,\label{infkey2}
\end{eqnarray}
can be obtained. A vacuum case is not compatible since equation (\ref{infkeya}) gives $S'=0$, which contradicts equation (\ref{inf11a}). 

\subsubsection{EOS (I) and (II)}
If a perfect fluid obeys EOS (I), we find from equations (\ref{infmu}) and (\ref{infp}) that

\begin{eqnarray}
P_1=W_1=0,\quad P_2 =K(8\pi G)^{1-\gamma}W_2^\gamma. \quad \mbox{[case (A)]}
\end{eqnarray}
If a perfect fluid obeys EOS (II), we find from equations (\ref{infmu}) and (\ref{infp}) that

\begin{eqnarray}
P_1=W_1=0,\quad P_2 =\frac{K}{(8\pi G)^{\gamma-1} m_b^{\gamma}}\left(W_2-\frac{P_2}{\gamma-1}\right)^\gamma. \quad \mbox{[case (B)]}
\end{eqnarray}
If $P_1=W_1=0$, equation (\ref{infkey2}) implies that these are included in the case of $p=K\mu$, which will be treated in the next subsection.

\subsubsection{EOS (III)}
If a perfect fluid obeys EOS (III), we find from equations (\ref{infmu}) and (\ref{infp}) that 

\begin{eqnarray}
P_1=K W_1, \quad P_2=KW_2. \quad \mbox{[case (C)]}\label{casec2} 
\end{eqnarray}
In this case, equation (\ref{infkey2}) implies that $K=-1$ or $P_2=W_2=0$. For $K=-1$, equations result in 

\begin{eqnarray}
P_1&=&W_1=0,\\
P_2&=&-W_2=-\frac{1}{s_0^2},\\
M_1&=&0,\\
M_2 &=&S=s_0,
\end{eqnarray}
where $s_0$ is a constant. This constant satisfies $1=K(8\pi G)^{1-\gamma}s_0^{-2(\gamma-1)}$ for case (A), while $-1=K/[(8\pi G)^{\gamma-1} m_b^{\gamma}][\gamma/(\gamma-1)]^{\gamma}s_0^{-2(\gamma-1)}$ for case (B). The metric functions $\Phi$ and $\Psi$ are obtained by the integration of the following differential equations:

\begin{eqnarray}
\Phi''&=&-\frac{1}{s_0^2}e^{2\Psi}-\Phi'^2+\Phi'\Psi', \label{lambdaeq1} \\
\Psi''&=&(\Phi'-\Psi'-1)\Psi'.\label{lambdaeq2}
\end{eqnarray}
In this solution, the perfect fluid is a cosmological constant since $p+\mu=0$. There is no physical center because of the constant circumferential radius. Although we have not obtained the analytic form of general solutions of equations (\ref{lambdaeq1}) and (\ref{lambdaeq2}), there is a special solution in which $\Psi$ is constant. For this case, the above equations are explicitly integrated as

\begin{eqnarray}
e^{\Phi}&=&A \sin\left(\frac{c_1}{s_0}\ln\xi\right)+B \cos\left(\frac{c_1}{s_0}\ln\xi\right), \\
e^{\Psi}&=&c_1,
\end{eqnarray}
where $A$, $B$ and $c_1$ are constants. With coordinate transformations $t^{c_1/s_0}=t'$ and $r^{c_1/s_0}=r'$, the solution in another form

\begin{equation}
ds^2=-\frac{s_0^2}{c_1^2}t'^{2s_0/c_1-2}\left[A \sin\left(\ln r'-\ln t'\right)+B \cos\left(\ln r'-\ln t'\right)\right]^2dt'^2+\frac{s_0^2}{r'^2}(dr'^2+r'^2d\Omega^2), \label{nariailike1}
\end{equation}
\begin{eqnarray}
2Gm&=&s_0,\\
8\pi G p&=&-8\pi G\mu=-\frac{1}{s_0^2},\label{nariailike2}
\end{eqnarray}
is obtained. We call this solution $\Lambda$-cylinder solution. It is clear that this new solution is not identical with but closely related to the solution found by Nariai~\cite{nariai}.

For $K \ne -1$ and $P_2=W_2 = 0$, equation (\ref{inf7}), (\ref{inf9}) and (\ref{inf11}) give

\begin{eqnarray}
\frac{W_1'}{W_1}=-\frac{2}{1+K}-\frac{S''}{S'}-2\frac{S'}{S}.\label{infkey}
\end{eqnarray}
It is noted that $S'$ cannot be zero since $S'=0$ contradicts equation (\ref{inf11a}). Substituting equations (\ref{inf00a}) and (\ref{inf11b}) for $P_1=KW_1$, and eliminating $\Phi', \Psi'$ and $W_1'/W_1$ with using equations (\ref{inf7}), (\ref{inf9}) and (\ref{infkey}), then we obtain the evolution equation for $S$:

\begin{eqnarray}
2(1+K)y'+2(1-K)y+3(1-K)(1+K)y^2=0,\label{evolS}
\end{eqnarray}
where $y \equiv S'/S$. From equations (\ref{evolS}) and (\ref{inf00a})-(\ref{inf11b}), we obtain

\begin{eqnarray}
e^{2\Phi}&=&-\frac{2}{(1+K)W_1}\frac{S'}{S},\label{dynab1st}\\
e^{2\Psi}&=&\frac{2K}{1+K}SS'+(1+3K)S'^2.
\end{eqnarray}
$W_1$, $M_1$ and $M_2$ can be obtained from equation (\ref{infkey}), (\ref{inf5}) and (\ref{inf6}), respectively:

\begin{eqnarray}
W_1&=&\frac{P_1}{K}=w_0\xi^{-2/(1+K)}S'^{-1}S^{-2},\\
M_1&=&e^{-2\Phi}SS'^2,\\
M_2&=&m_2=S(1-e^{-2\Psi}S'^2),
\end{eqnarray}
where $w_0$ and $m_2$ are constants. For $K \ne 1$, equation (\ref{evolS}) can be integrated to give

\begin{eqnarray}
S=s_1|1-s_0 \xi^{-\frac{1-K}{1+K}}|^{\frac{2}{3(1-K)}},\label{dynablast}
\end{eqnarray}
where $s_0$ and $s_1$ are integration constants. Since equation (\ref{dynab1st}) implies $W_1\propto \xi^{-2/(1+K)}S^{-2} S'^{-1}\propto \xi^{-1}|1-s_0 \xi^{-(1-K)/(1+K)}|^{-(1+K)/(1-K)}$, this solution is singular at the physical center $S=0$. For $s_0>0$, this solution represents collapsing shells, since $S$ decreases as $\xi$ decreases from $\xi=+\infty$ which means that $t$ increases from $t=+0$ for constant $r$. A shell for some constant $r$ collapses to a singularity $S=0$ at a finite time $t_1=s_0^{-(1+K)/(1-K)}r$. For $s_0<0$, this solution represents expanding shells, since $S$ increases as $\xi$ decreases from $\xi=+\infty$ which means that $t$ increases from $t=+0$ for constant $r$. In this case, $S$ cannot be zero. We call this the dynamical solution (B). For $K=1$, equation (\ref{evolS}) can be integrated to give the flat FRW solution: 

\begin{eqnarray}
e^{\Phi}&=&c_0,\\
e^{\Psi}&=&\frac13 s_0 \xi^{-1/3},\\
S&=&s_0 \xi^{-1/3},\\
M_1&=&\frac19 c_0^{-2}s_0^3 \xi^{-1},\\
M_2&=&0,\\
P_1&=&W_1=\frac13 c_0^{-2},\\
P_2&=&W_2=0,
\end{eqnarray}
where $c_0$ and $s_0$ are constants. With the coordinate transformations $t'=c_0 t$ and $r'=(s_0/c_0^{1/3}) r^{-1/3}$, the solution in a more usual form 

\begin{eqnarray}
ds^2&=&-dt'^2+t'^{2/3}(dr'^2+r'^2d\Omega^2),\\
2Gm&=&\frac19 \frac{r'}{t'},\\
8\pi G p&=&8\pi G \mu=\frac{1}{3t'^2},
\end{eqnarray}
is obtained.

\section{$\xi^{\mu}$ parallel to $U^{\mu}$}
\label{sec:parallel}

\subsection{Self-similarity of the second kind}
In this case, we can choose ${\bf \xi}$ to be 

\begin{eqnarray}
\xi^{\mu}\frac{\partial}{\partial x^{\mu}}=t\frac{\partial}{\partial t},
\end{eqnarray}
and then the metric can then be written as
\begin{eqnarray}
ds^2 &=& -t^{2(\alpha-1)}e^{2\Phi(r)}dt^2+t^2dr^2+S(r)^2t^2d\Omega^2.\label{linepara}
\end{eqnarray}
In this case, the Einstein equations imply that the quantities $m, \mu$ and $p$ must be of the form 

\begin{eqnarray}
2Gm&=&tM_1(r)+t^{3-2\alpha}M_2(r),\label{param}\\
8\pi G \mu&=&t^{-2}W_1(r)+t^{-2\alpha}W_2(r),\label{paramu}\\
8\pi G p&=&t^{-2}P_1(r)+t^{-2\alpha}P_2(r),\label{parap}
\end{eqnarray}
and a set of ordinary differential equations is obtained when one demands that the Einstein equations and the equations of motion for the matter field are satisfied for the $O[t^0]$ and $O[t^{2-2\alpha}]$ terms separately. The resulting equations for a perfect fluid (\ref{basic1})-(\ref{11}) reduce to the following:

\begin{eqnarray}
M_1&=&-P_1S^3,\label{para1}\\
(3-2\alpha)M_2&=&-P_2S^3,\label{para2}\\
M_1'&=&W_1S^2S',\label{para3}\\
M_2'&=&W_2S^2S',\label{para4}\\
M_1&=&S[1-S'^2],\label{para5}\\
M_2&=&S^3 e^{-2\Phi},\label{para6}\\
3P_1&=&-W_1,\label{para7}\\
3P_2&=&(2\alpha-3) W_2,\label{para8}\\
-P_1'&=&(P_1+W_1)\Phi',\label{para9}\\
-P_2'&=&(P_2+W_2)\Phi',\label{para10}\\
0&=&S\Phi',\label{para11}\\
2S''S+S'^2&=&1-W_1S^2, \label{para00a}\\
3&=&W_2 e^{2\Phi}, \label{para00b}\\
S'^2+2\Phi'S'S&=&1+P_1S^2, \label{para11a}\\
2\alpha-3&=&P_2 e^{2\Phi}, \label{para11b}
\end{eqnarray}
where the prime denotes the derivative with respect to $r$. A vacuum spacetime is not compatible with this case since $W_2=0$ contradicts equation (\ref{para00b}).

\subsubsection{EOS (I) and (II)}
If a perfect fluid obeys EOS (I), we find from equations (\ref{paramu}) and (\ref{parap}) that

\begin{eqnarray}
\alpha=\gamma, \quad P_1=W_2=0,\quad P_2 =\frac{K}{(8\pi G)^{\gamma-1}}W_1^{\gamma}, \quad \mbox{[case (A)]}
\end{eqnarray}
or
\begin{eqnarray}
\alpha=\frac{1}{\gamma}, \quad P_2=W_1=0,\quad P_1 =\frac{K}{(8\pi G)^{\gamma-1}}W_2^{\gamma}. \quad \mbox{[case (B)]}
\end{eqnarray}
If a perfect fluid obeys EOS (II), we find from equations (\ref{paramu}) and (\ref{parap}) that

\begin{eqnarray}
\alpha=\gamma,P_1=0,\quad P_2 =\frac{K}{(8\pi G)^{\gamma-1} m_b^{\gamma}}W_1^{\gamma}=(\gamma-1)W_2, \quad \mbox{[case (C)]}
\end{eqnarray}
or
\begin{eqnarray}
\alpha=\frac{1}{\gamma},P_2=0,\quad P_1 =\frac{K}{(8\pi G)^{\gamma-1} m_b^{\gamma}}W_2^{\gamma}=(\gamma-1)W_1. \quad \mbox{[case (D)]}
\end{eqnarray}

Since $W_2$ cannot be zero, case (A) is excluded. If $P_1=0$, equation (\ref{para7}) gives $W_1=0$, which implies $W_2=0$ for case (C). Hence case (C) is also excluded.

For $P_2=0$, equation (\ref{para11b}) gives $\alpha=3/2$ which implies $\gamma=2/3$ for both cases (B) and (D), and then equations reduce to the following:

\begin{eqnarray}
e^{\Phi}&=&c_0,\\
P_1&=&-\frac13 W_1=p_0,\\
W_2&=&\frac{3}{c_0^2},\\
M_1&=&-P_1 S^3,\\
M_2&=&c_0^{-2}S^3,\\
-p_0 S^2&=& 1-S'^2,
\end{eqnarray}
where $c_0$ and $p_0$ are constants. $P_1=-W_1/3$ implies that the spacetime must be vacuum in case (B). We require that $W_1$ is positive so that $p_0<0$. The solution is 

\begin{eqnarray}
S=\frac{1}{\sqrt{-p_0}}\sin{\sqrt{-p_0}r},
\end{eqnarray}
which gives the metric 

\begin{eqnarray}
ds^2 = -c_0^2 t dt^2+t^2\left(dr^2+\frac{1}{(-p_0)}\sin^2\sqrt{-p_0}r d\Omega^2\right).
\end{eqnarray}
With the coordinate transformations $t'=(2c_0/3)t^{3/2}$ and $r'=\sqrt{-p_0}r$, this solution is found to be the closed FRW solution with dust and $p=-\mu/3$ comoving fluids: 

\begin{eqnarray}
ds^2 &=& -dt'^2+(-p_0)^{-1}\left(\frac{3}{2c_0}\right)^{4/3}t'^{4/3}\left(dr'^2+\sin^2r' d\Omega^2\right),\\
2Gm&=&\left[-\left(\frac32\right)^{\frac23}c_0^{-2/3}(-p_0)^{-1/2}t'^{2/3}+c_0^{-2}(-p_0)^{-3/2}\right]\sin^3r',\\
8\pi G p&=&\left(\frac{3}{2c_0}\right)^{-4/3}p_0t'^{-4/3},\\
8\pi G \mu&=&-3p_0 \left(\frac{3}{2c_0}\right)^{-4/3}t'^{-4/3}+\frac43 t'^{-2}.
\end{eqnarray}

\subsubsection{EOS (III)}
If a perfect fluid obeys EOS (III), we find from equations (\ref{paramu}) and (\ref{parap}) that 

\begin{eqnarray}
P_1=K W_1, \quad P_2=KW_2. \quad \mbox{[case (E)]}
\end{eqnarray}
In this case, equations (\ref{para7}) and (\ref{para8}) imply that $\alpha=3/2$ or $W_1=0$. $\alpha=3/2$ implies $P_2=W_2=0$ from equation (\ref{para8}) or (\ref{para11b}). $W_2=0$ contradicts equation (\ref{para00b}). Hence $W_1$ must be zero. The resulting solution is

\begin{eqnarray}
ds^2&=&-t^{2(\alpha-1)}c_0^2dt^2+t^2(dr^2+r^2d\Omega^2),\\
P_1&=&W_1=0,\\
P_2&=&\left(\frac23\alpha-1\right)W_2=(2\alpha-3)c_0^{-2},\\
M_1&=&0,\\
M_2&=&c_0^{-2} r^3,
\end{eqnarray}
where $c_0$ is a constant. This is the flat FRW solution. With the coordinate transformations $(c_0/\alpha) t^{\alpha}= t'$ and $(\alpha/c_0)^{1/\alpha}r=r'$, the solution in a more usual form 

\begin{eqnarray}
ds^2&=&-dt'^2+t'^{\frac{2}{\alpha}}(dr'^2+r'^2d\Omega^2),\\
2Gm&=&\alpha^{-\frac{3}{\alpha}}r'^3,\\
8\pi G p&=&8\pi G \left(\frac23\alpha-1\right)\mu=\frac{2\alpha-3}{\alpha^2}t'^{-2},
\end{eqnarray}
is obtained. This solution exists for $-1< K\le 1~(K \ne 0,-1/3)$ which implies $0< \alpha \le 3~(\alpha \ne 3/2,1)$.

\subsection{Self-similarity of the zeroth kind}
Equations (\ref{linepara})-(\ref{para11}) can be used for this case with $\alpha=0$. A vacuum spacetime is also not compatible with this case.

\subsubsection{EOS (I) and (II)}

If a perfect fluid obeys EOS (I), we find from equations (\ref{paramu}) and (\ref{parap}) that

\begin{eqnarray}
P_1=W_1=0,\quad P_2 =K(8 \pi G)^{1-\gamma}W_2^\gamma. \quad \mbox{[case (A)]}
\end{eqnarray}
If a perfect fluid obeys EOS (II), we find from equations (\ref{paramu}) and (\ref{parap}) that

\begin{eqnarray}
P_1=W_1=0,\quad P_2 =\frac{K}{(8\pi G)^{\gamma-1} m_b^{\gamma}}\left(W_2-\frac{P_2}{\gamma-1}\right)^\gamma. \quad \mbox{[case (B)]}
\end{eqnarray}

Equation (\ref{para8}) gives $P_2+W_2=0$ so that above two cases are included in the case of $p=K\mu$, which will be treated in the next subsection.

\subsubsection{EOS (III)}
If a perfect fluid obeys EOS (III), we find from equations (\ref{paramu}) and (\ref{parap}) that 

\begin{eqnarray}
P_1=K W_1, \quad P_2=KW_2. \quad \mbox{[case (C)]}
\end{eqnarray}
In this case, equations (\ref{para7}) and (\ref{para8}) imply that $W_1=P_1=0$ since $W_2$ cannot be zero. The resulting solution is 

\begin{eqnarray}
ds^2&=&-\frac{3}{(-p_0)}t^{-2}dt^2+t^2(dr^2+r^2d\Omega^2),\\
P_1&=&W_1=0,\\
P_2&=&-W_2=p_0,\\
M_1&=&0,\\
M_2&=&\frac{-p_0}{3} r^3,
\end{eqnarray}
where $p_0$ is a constant. This is the de-Sitter solution. With the coordinate transformation $t=\exp(\sqrt{-p_0/3}t')$, the solution in a more usual form 

\begin{eqnarray}
ds^2&=&-dt'^2+e^{2\sqrt{-p_0/3}t'}(dr^2+r^2d\Omega^2),\\
2Gm&=&\frac{-p_0}{3}r^3e^{3\sqrt{-p_0/3}t'},\\
8\pi G p&=&-8\pi G \mu=8\pi G p_0,
\end{eqnarray}
is obtained.

\subsection{Self-similarity of the infinite kind}

In this case, we can choose ${\bf \xi}$ to be 

\begin{eqnarray}
\xi^{\mu}\frac{\partial}{\partial x^{\mu}}=t\frac{\partial}{\partial t},
\end{eqnarray}
and then the metric can then be written as
\begin{eqnarray}
ds^2 &=& -e^{2\Phi(r)}dt^2+dr^2+S(r)^2d\Omega^2.
\end{eqnarray}
In this case, the Einstein equations imply that the quantities $m, \mu$ and $p$ must be of the form 

\begin{eqnarray}
2Gm&=&M(r),\\
8\pi G \mu&=&W(r),\\
8\pi G p&=&P(r),
\end{eqnarray}
and the Einstein equations and the equations of motion for the matter field (\ref{basic1})-(\ref{11}) are written as

\begin{eqnarray}
M&=&S(1-S'^2),\label{parainf1}\\
M'&=&WS'S^2,\label{parainf2}\\
(P+W)\Phi'&=&-P',\label{parainf3}\\
2SS''+S'^2&=&1-WS^2,\label{parainf00}\\
2\Phi'S'S+S'^2&=&1+PS^2,\label{parainf11}
\end{eqnarray}
where the prime denotes the derivative with respect to $r$. These equations give the Tolman-Oppenheimer-Volkoff equation. It implies that any spherically symmetric static spacetime is a self-similar solution of the infinite kind in which kinematic self-similar vector is parallel to the fluid flow. In a vacuum case, the Schwarzschild solution can be obtained as

\begin{eqnarray}
M&=&m_0,\\
r&=&\pm \sqrt{(S-m_0)S}\pm \frac{m_0}{2}\ln\left(S-\frac{m_0}{2}+\sqrt{(S-m_0)S}\right)+c_1,\\
e^{\Phi}&=&c_0\left(1-\frac{m_0}{S}\right)^{1/2},
\end{eqnarray}
where $c_0, c_1$ and $m_0$ are constants, while the Minkowski spacetime can be obtained as

\begin{eqnarray}
M&=&0, \quad S=r, \quad e^{\Phi}=c_0,
\end{eqnarray}
where $c_0$ is a constant.

\section{$\xi^{\mu}$ orthogonal to $U^{\mu}$}
\label{sec:orthogonal}

\subsection{Self-similarity of the second kind}
In this case, we can choose ${\bf \xi}$ to be 

\begin{eqnarray}
\xi^{\mu}\frac{\partial}{\partial x^{\mu}}=r\frac{\partial}{\partial r},
\end{eqnarray}
and then the metric can then be written as
\begin{eqnarray}
ds^2 &=& -r^{2\alpha}dt^2+e^{2\Psi(t)}dr^2+S(t)^2r^2d\Omega^2.\label{line}
\end{eqnarray}
In this case, the Einstein equations imply that the quantities $m, \mu$ and $p$ must be of the form 

\begin{eqnarray}
2Gm&=&rM_1(t)+r^{3-2\alpha}M_2(t),\label{orthm}\\
8\pi G \mu&=&r^{-2}W_1(t)+r^{-2\alpha}W_2(t),\label{orthmu}\\
8\pi G p&=&r^{-2}P_1(t)+r^{-2\alpha}P_2(t),\label{orthp}
\end{eqnarray}
and a set of ordinary differential equations is obtained when one demands that the Einstein equations and the equations of motion for the matter field are satisfied for the $O[r^0]$ and $O[r^{2-2\alpha}]$ terms separately. It is noted that the solution is always singular at $r=0$, which is correspond to the physical center. The equations for a perfect fluid (\ref{basic1})-(\ref{22}) reduce to the following:

\begin{eqnarray}
M_1&=&W_1S^3,\label{orth1}\\
(3-2\alpha)M_2&=&W_2S^3,\label{orth2}\\
M_1'&=&-P_1S^2S',\label{orth3}\\
M_2'&=&-P_2S^2S',\label{orth4}\\
M_1&=&S(1-e^{-2\Psi}S^2),\label{orth5}\\
M_2&=&SS'^2 ,\label{orth6}\\
(2-\alpha)P_1&=&\alpha W_1,\label{orth7}\\
\alpha P_2&=&\alpha W_2,\label{orth8}\\
W_1'S&=&-(P_1+W_1)(\Psi'S+2S'),\label{orth9}\\
W_2'S&=&-(P_2+W_2)(\Psi'S+2S'),\label{orth10}\\
(1-\alpha)S'&=&S\Psi',\label{orth11}\\
-W_1S^2&=&e^{-2\Psi}S^2-1,\label{orth00a}\\
W_2S^2&=&S'^2+2\Psi'S'S,\label{orth00b}\\
-P_1S^2&=&1-(1+2\alpha)S^2 e^{-2\Psi},\label{orth11a}\\
-P_2S^2&=&2S''S+S'^2,\label{orth11b}\\
P_1&=&\alpha^2 e^{-2\Psi},\label{orth22a}\\
-P_2 S&=&S''+\Psi'S'+\Psi''S+\Psi'^2 S,\label{orth22b}
\end{eqnarray}
where the prime denotes the derivative with respect to $t$. A vacuum spacetime is not compatible with this case since $P_1=0$ contradicts equation (\ref{orth22a}).

\subsubsection{EOS (I) and (II)}
If a perfect fluid obeys EOS (I), we find from equations (\ref{orthmu}) and (\ref{orthp}) that

\begin{eqnarray}
\alpha=\gamma, \quad P_1=W_2=0,\quad P_2 =\frac{K}{(8\pi G)^{\gamma-1}}W_1^{\gamma}, \quad \mbox{[case (A)]}
\end{eqnarray}
or
\begin{eqnarray}
\alpha=\frac{1}{\gamma}, \quad P_2=W_1=0,\quad P_1 =\frac{K}{(8\pi G)^{\gamma-1}}W_2^{\gamma}. \quad \mbox{[case (B)]}
\end{eqnarray}
If a perfect fluid obeys EOS (II), we find from equations (\ref{orthmu}) and (\ref{orthp}) that

\begin{eqnarray}
\alpha=\gamma,P_1=0,\quad P_2 =\frac{K}{(8\pi G)^{\gamma-1} m_b^{\gamma}}W_1^{\gamma}=(\gamma-1)W_2, \quad \mbox{[case (C)]}\label{casec2}
\end{eqnarray}
or
\begin{eqnarray}
\alpha=\frac{1}{\gamma},P_2=0,\quad P_1 =\frac{K}{(8\pi G)^{\gamma-1} m_b^{\gamma}}W_2^{\gamma}=(\gamma-1)W_1. \quad \mbox{[case (D)]}\label{cased2}
\end{eqnarray}

Cases (A) and (C) are excluded since $P_1=0$ contradicts equation (\ref{orth22a}). For $P_2=0$, equation (\ref{orth8}) gives $W_2=0$, which results in $P_1=0$ for both cases (B) and (D), and therefore it can be concluded that there are no solutions in these cases.

\subsubsection{EOS (III)}
If a perfect fluid obeys EOS (III), we find from equations (\ref{orthmu}) and (\ref{orthp}) that 

\begin{eqnarray}
P_1=K W_1, \quad P_2=KW_2 \quad \mbox{[case (E)]}.
\end{eqnarray}
$W_1W_2=0$ is concluded, since if not, equation (\ref{orth7}) contradicts equation (\ref{orth8}). Equation (\ref{orth22a}) implies that $P_1$ cannot be zero, and hence $W_2 \ne 0$ so that $W_2=P_2=0$ can be concluded. Then equations (\ref{orth2}) and (\ref{orth4}) result in $(3-2\alpha)M_2=0$ where $M_2$ is constant. Equation (\ref{orth7}) implies that $K=\alpha/(2-\alpha)$, and then $-1 \le K \le 1$ implies that $\alpha$ cannot be $3/2$. Hence $M_2=0$ is concluded. The resulting solution is 

\begin{eqnarray}
ds^2&=&-r^{2\alpha}dt^2+\frac{s_0^2}{1-w_0 s_0^2}dr^2+s_0^2 r^2 d\Omega^2,\label{static1perp}\\
M_1&=&w_0 s_0^3, \quad M_2=0,\\
P_1&=&\frac{\alpha}{2-\alpha}W_1=\frac{\alpha}{2-\alpha}w_0,\\
P_2&=&W_2=0,\label{static2perp}
\end{eqnarray}
where $w_0$ and $s_0$ are constants and $(1+2\alpha-\alpha^2)w_0s_0^2=\alpha(2-\alpha)$. With the coordinate transformations $s_0^{-\alpha}t=t'$ and $s_0 r=r'$, this solution can be written 

\begin{eqnarray}
ds^2&=&-r'^{2\alpha}dt'^2+\frac{1}{1-w_0 s_0^2}dr'^2+r'^2 d\Omega^2,\\
2Gm&=&w_0s_0^2r',\\
8\pi G p&=&8\pi G\frac{\alpha}{2-\alpha}\mu=\frac{\alpha}{2-\alpha}\frac{w_0s_0^2}{r'^2}.
\end{eqnarray}
This solution represents the same singular static solution as equations (\ref{static1})-(\ref{static2}) with $K=\alpha/(2-\alpha)$. $K=\alpha/(2-\alpha)$ gives $K \ne -1$ and the assumption $\alpha \ne 1 $ corresponds to $K \ne 1$. The positive energy density requires $-1<K <2\sqrt2-3$ and $0 < K <1$ which correspond to $\alpha< (2\sqrt2-3)/(\sqrt2-1)$ and $0<\alpha<1$.

\subsection{Self-similarity of the zeroth kind}
Equations (\ref{orth1})-(\ref{orth22b}) can be used for this case with $\alpha=0$. In a vacuum case, the Minkowski spacetime can be obtained since $M_1=M_2=0$.

\subsubsection{EOS (I) and (II)}
If a perfect fluid obeys EOS (I), we find from equations (\ref{orthmu}) and (\ref{orthp}) that

\begin{eqnarray}
P_1=W_1=0,\quad P_2 =K(8 \pi G)^{1-\gamma}W_2^\gamma. \quad \mbox{[case (A)]}\label{casea3}
\end{eqnarray}
If a perfect fluid obeys EOS (II), we find from equations (\ref{orthmu}) and (\ref{orthp}) that

\begin{eqnarray}
P_1=W_1=0,\quad P_2 =\frac{K}{(8\pi G)^{\gamma-1} m_b^{\gamma}}\left(W_2-\frac{P_2}{\gamma-1}\right)^\gamma. \quad \mbox{[case (B)]}\label{caseb3}
\end{eqnarray}

In this case, equations reduce to 

\begin{eqnarray}
M_1&=&0,\\
M_2&=&\frac13 W_2 S^3,\\
e^{2\Psi}&=&S^2,\label{special1}\\
W_2S^2&=&S'^2+2\Psi'S'S,\\
-P_2S^2&=&2S''S+S'^2\label{special2},
\end{eqnarray}
which give the flat FRW metric 

\begin{eqnarray}
ds^2=-dt^2+S^2(dr^2+r^2d\Omega^2).
\end{eqnarray}
The evolution for $S$ is governed by

\begin{eqnarray}
W_2S^2&=&3S'^2,\label{frw1}\\
-P_2S^2&=&2S''S+S'^2.\label{frw2}
\end{eqnarray}
These equations reduce to a evolution equation for $S$ by using of EOS (I) or (II). $S$ is generally not a power-law function of $t$ in both cases.

\subsubsection{EOS (III)}
If a perfect fluid obeys EOS (III), we find from equations (\ref{orthmu}) and (\ref{orthp}) that 

\begin{eqnarray}
P_1=K W_1, \quad P_2=KW_2. \quad \mbox{[case (C)]}\label{casec3} 
\end{eqnarray}
In this case, equation (\ref{orth7}) shows $P_1=0$, and hence $W_1 = 0$. Equations (\ref{frw1}) and (\ref{frw2}) can be integrated to give $S=s_0 t^{\frac{2}{3(1+K)}}$ where $s_0$ is a constant, which is the flat FRW spacetime for $K \ne -1$, while $S=s_0 e^{\sqrt{(-p_0)/3}t}~(p_0 \equiv P_2=-W_2<0)$, which is the de-Sitter spacetime for $K=-1$. With the coordinate transformation $s_0 r=r'$, a more usual form of the flat FRW solution: 

\begin{eqnarray}
ds^2&=&-dt^2+t^{\frac{2}{3(1+K)}}(dr'^2+r'^2 d\Omega^2),\\
2Gm&=&\frac{4}{9(1+K)^2}t^{-\frac{2K}{1+K}}r'^3,\\
8\pi G p&=&8\pi GK\mu=\frac{4}{3(1+K)^2}\frac{1}{t^2},
\end{eqnarray}
and the de-Sitter solution:

\begin{eqnarray}
ds^2&=&-dt^2+e^{2\sqrt{(-p_0)/3}t}(dr'^2+r'^2 d\Omega^2),\\
2Gm&=&\frac{(-p_0)}{3}e^{3\sqrt{(-p_0)/3}t}r'^3,\\
8\pi G p&=&-8\pi G\mu=p_0,
\end{eqnarray}
are obtained.

\subsection{Self-similarity of the infinite kind}

In this case, we can choose ${\bf \xi}$ to be 

\begin{eqnarray}
\xi^{\mu}\frac{\partial}{\partial x^{\mu}}=\frac{\partial}{\partial r},
\end{eqnarray}
and then the metric can then be written as
\begin{eqnarray}
ds^2 &=& -e^{2r}dt^2+e^{2\Psi(t)}dr^2+S(t)^2d\Omega^2.\label{line}
\end{eqnarray}
In this case, the Einstein equations imply that the quantities $m, \mu$ and $p$ must be of the form 

\begin{eqnarray}
2Gm&=&e^{-2r}M_1(t)+M_2(t),\label{orthinfm}\\
8\pi G \mu&=&e^{-2r}W_1(t)+W_2(t),\label{orthinfmu}\\
8\pi G p&=&e^{-2r}P_1(t)+P_2(t).\label{orthinfp}
\end{eqnarray}
A set of ordinary differential equations is obtained when one demands that the Einstein equations and the equations of motion for the matter field are satisfied for the $O[r^0]$ and $O[e^{-2r}]$ terms separately. The resulting equations for a perfect fluid (\ref{basic1})-(\ref{11}) are

\begin{eqnarray}
S&=&M_2=s_0,\\
M_1&=&0,\\
P_1&=&W_1=0,\\
P_2&=&-W_2=-w_0=-\frac{1}{s_0^2}.
\end{eqnarray}
where $s_0$ and $w_0$ are constants. Equation (\ref{22}) gives $e^{2\Psi}=P_2=-1/w_0$. Since $e^{2\Psi}S^{-2}=-1<0$, it can be concluded that there are no solutions in this case, independent of the form of the equation of state. 

\section{Summary and Discussions}
\label{sec:summary}

We have classified the kinematic self-similar perfect-fluid solutions with either EOS (I), (II) or (III). In most cases, the governing equations can be integrated to give exact solutions, although there are a few exceptions. The analytic form of general solutions has not been obtained in the infinite-kind case with EOS (III) for $K=-1$ in which a kinematic self-similar vector is tilted. The results are summarized in table \ref{tablesol1}-\ref{tablesol4}. We note that, independent of the form of the equation of state, any static solution is a kinematic self-similar solution of the infinite kind in the parallel case, since the governing equations give the Tolman-Oppenheimer-Volkov equation. It should also be noted that, independent of the form of the equation of state, kinematic self-similarity of the infinite kind in the orthogonal case is incompatible with a spherically symmetric spacetime. 

In the cases of EOS (I) and (II), i.e., the polytropic equation of state, the FRW solution is one of the compatible solutions. The closed FRW solution with dust and $p=-\mu/3$ comoving fluids is the second-kind solution in the parallel case for EOS (II) with $\gamma=2/3$, while the flat FRW solution is the zeroth-kind solution in the orthogonal case for both EOS (I) and (II), in which the scale factor is not a power-law function of $t$ in general. 

Next we summarize the case of EOS (III). The flat FRW solution is the second-kind solution in the tilted case for $\alpha \ne 3(1+K)/2$ and in the parallel case for $\alpha = 3(1+K)/2$, the zeroth-kind solution in the orthogonal case for $-1< K \le 1$ and $K \ne 0$ and the infinite-kind solution in the tilted case for $K =1$. The de-Sitter solution is the zeroth-kind solution both in the tilted, parallel and orthogonal cases for $K=-1$. The singular static solution, which is singular at the physical center, is both the second-kind solution for any $\alpha$ and zeroth-kind solution in the tilted case. Because of its staticity, this solution is also the infinite-kind solution in the parallel case. Positivity of energy density of the perfect fluid requires that $-1<K<2\sqrt2-3$ or $0<K \le 1$ in these cases. The singular static solution is also the second-kind solution in the orthogonal case for $K=\alpha/(2-\alpha)$. Positivity of energy density implies that $-1<K<2\sqrt2-3$ or $0<K < 1$ in this case. We have found two interesting exact solutions, which we call the dynamical solutions (A) and (B), respectively. The dynamical solution (A) is the second-kind solution in the tilted case for $\alpha=3/2$ and $K \ne -1$, while the dynamical solution (B) is the infinite-kind solution in the tilted case for $K \ne -1,1$. These solutions represent collapsing or expanding shells. We have also found another new exact solution, which we call $\Lambda$-cylinder solution. $\Lambda$-cylinder solution is the infinite-kind solution in the tilted case for $K=-1$ and describes a spacetime with constant circumferential radius and no physical center. It should be noted that we have not obtained the analytic form of general solutions in the infinite-kind case for $K=-1$ in which a kinematic self-similar vector is tilted. The dynamical solutions (A) and (B) and $\Lambda$-cylinder solution could be important and we are now researching on these solutions~\cite{mhio}. 

In the vacuum case, the Schwarzschild solution is the second-kind solution in the tilted case for $\alpha=3/2$ and infinite-kind solution in the parallel case, while the Minkowski spacetime has kinematic self-similarity of the second kind in the tilted case for any $\alpha$, of the zeroth kind in both the tilted and orthogonal cases and of the infinite kind in the parallel case.

The polytropic perfect-fluid solutions compatible with kinematic self-similarity are the FRW solution and general static solutions. This result differs from that in Newtonian case, for which there exists the polytropic Larson-Penston solution which is a self-similar solution describing a collapsing polytrope gas~\cite{hm2000a,yahil1983,ss1988}. The result in general relativity depends on the fact that $P_1=0$ or $P_2=0$. The non-trivial solutions could arise when we assume the equation of state in which it is possible to have $P_1 P_2 \ne 0$. We discuss the possible equations of state which are compatible with self-similarity in appendix~\ref{sec:possibleeos}. A more general kind of self-similarity called {\it partial homothety}, which imposes only equation (\ref{kss}) on a spacetime, could be compatible with EOS (I) or (II)~\cite{ph}. Such cases will be investigated elsewhere. 

If the polytropic Larson-Penston solution is an attractor of generic non-self-similar collapse of polytropic gas in Newtonian gravity, as for the isothermal case, then how does the collapse proceed in the relativistic regime when we assume that the polytropic equation of state in relativistic regime is described by EOS (I) or (II)? For $0<\gamma<1$, both EOS (I) and (II) are approximated by a dust fluid so that the generic collapse could converge to the spacetime whose central region can be described by the Tolman-Bondi solution~\cite{LTB}. For $1<\gamma$, EOS (II) is approximated by $p=(\gamma-1)\mu$. For $1<\gamma \lesssim 1.036$, there exists the general relativistic Larson-Penston solution, which is the self-similar solution of the first kind. Generic collapse, in the neighborhood of the center, could converge to this solution in an approach to a singularity. However, for $1.036 \lesssim \gamma$, the attractor solution loses its attractive nature. For $4/3 <\gamma<2$, the collapsing flat Friedmann solution, which is also a self-similar solution of the first kind, could be an attractor solution, since it has been shown by means of mode analyses that the solution has no unstable modes for spherical perturbations~\cite{harada2001}. Stable self-similar solutions of the first kind for $1.036 \lesssim \gamma \le 4/3$ have not been found so far. The candidate for an attractor solution is not known. For $2<\gamma$, the dominant energy condition could be violated in the central regime as the collapse proceeds, which is unphysical.

In order to understand the whole picture of the generic collapse of a polytrope gas which obeys the equation of state (I) or (II), full numerical simulations of gravitational collapse will be quite helpful. 

\begin{table}[htbp]
	\begin{center}
		\begin{tabular}{llccccc}
		Self-similarity & Solution  \\ 
		\noalign{\hrule height 0.8pt}
		1st (tilted) & none \\  \hline 
		1st ($\parallel$) & none  \\  \hline 
		1st ($\perp$) & none \\  \hline 
		2nd (tilted) & none \\  \hline 
		2nd ($\parallel$) & none  \\  \hline 
		2nd ($\perp$) & none \\  \hline 
		zeroth (tilted) & none \\  \hline 
		zeroth ($\parallel$) & none  \\ \hline 
		zeroth ($\perp$) & flat FRW  \\ \hline 
		infinite (tilted) & none  \\ \hline 
		infinite ($\parallel$) & All static solutions with the EOS (I)  \\ \hline
		infinite ($\perp$) & none  \\ 
		\noalign{\hrule height 0.8pt}
		\end{tabular}
	\end{center}
	\caption{Kinematic self-similar solutions for the EOS (I). $\parallel$ and $\perp$ denote the parallel and the orthogonal cases, respectively. See text for the values of parameters, $\alpha$, $K$ and $\gamma$.}
	\label{tablesol1}
\end{table}

\begin{table}[htbp]
	\begin{center}
		\begin{tabular}{llccccc}
		Self-similarity & Solution  \\ 
		\noalign{\hrule height 0.8pt}
		1st (tilted) & none \\  \hline 
		1st ($\parallel$) & none  \\  \hline 
		1st ($\perp$) & none \\  \hline 
		2nd (tilted) & none  \\  \hline 
		2nd ($\parallel$) & closed FRW  \\ \hline
		2nd ($\perp$) & none \\ \hline 
		zeroth (tilted) & none \\ \hline 
		zeroth ($\parallel$) & none \\ \hline 
		zeroth ($\perp$) & flat FRW  \\ \hline 
		infinite (tilted) & none \\ \hline 
		infinite ($\parallel$) & All static solutions with the EOS (II)   \\ \hline
		infinite ($\perp$) & none \\ 
		\noalign{\hrule height 0.8pt}
		\end{tabular}
	\end{center}
	\caption{Kinematic self-similar solutions for the EOS (II). $\parallel$ and $\perp$ denote the parallel and the orthogonal cases, respectively. See text for the values of parameters, $\alpha$, $K$ and $\gamma$.}
	\label{tablesol2}
\end{table}

\begin{table}[htbp]
	\begin{center}
		\begin{tabular}{llccccc}
		Self-similarity & Solution & Equation number \\ 
		\noalign{\hrule height 0.8pt}
		1st (tilted) & see Refs~\citen{cc2000,op1990} & \\  \hline 
		1st ($\parallel$) & FRW (see Ref~\citen{coley1991})  & \\  \hline 
		1st ($\perp$) & see Ref~\citen{mcintosh1975} & \\  \hline 
		2nd (tilted) & flat FRW & \\
		             & Singular static & (\ref{static1})-(\ref{static2}) \\
		             & Dynamical solution (A) & (\ref{dynaa1st}),~(\ref{dynaa2nd})-(\ref{dynaalast})\\ \hline
		2nd ($\parallel$) & flat FRW & \\ \hline
		2nd ($\perp$) & Singular static & (\ref{static1perp})-(\ref{static2perp}) \\ \hline 
		zeroth (tilted) & de-Sitter & \\ 
		             & Singular static & (\ref{static1zero})-(\ref{static2zero}) \\ \hline
		zeroth ($\parallel$) & de-Sitter & \\ \hline 
		zeroth ($\perp$) & flat FRW & \\ 
		             & de-Sitter & \\ \hline
		infinite (tilted) & flat FRW & \\
		             & Dynamical solution (B) & (\ref{dynab1st})-(\ref{dynablast})\\
		             & $\Lambda$-cylinder solution & (\ref{nariailike1})-(\ref{nariailike2}) \\ \hline
		infinite ($\parallel$) & All static solutions  with the EOS (III)&  \\ \hline
		infinite ($\perp$) & none & \\ 
		\noalign{\hrule height 0.8pt}
 
		\end{tabular}
	\end{center}
	\caption{Kinematic self-similar solutions for the EOS (III). $\parallel$ and $\perp$ denote the parallel and the orthogonal cases, respectively. See text for the values of parameters, $\alpha$, $K$. It is noted that the analytic form of general solutions has not been obtained in the infinite kind case with $K=-1$ in which a kinematic self-similar vector is tilted. }
	\label{tablesol3}
\end{table}

\begin{table}[htbp]
	\begin{center}
		\begin{tabular}{llccccc}
		Self-similarity & Solution  \\ 
		\noalign{\hrule height 0.8pt}
		2nd (tilted) & Minkowski  \\
		             & Schwarzschild  \\ \hline
		2nd ($\parallel$) & none \\ \hline 
		2nd ($\perp$) & none \\ \hline 
		zeroth (tilted) & Minkowski  \\ \hline 
		zeroth ($\parallel$) & none \\ \hline 
		zeroth ($\perp$) & Minkowski  \\ \hline 
		infinite (tilted) & none \\ \hline 
		infinite ($\parallel$) & Minkowski  \\
		             & Schwarzschild  \\ \hline
		infinite ($\perp$) & none \\ 
		\noalign{\hrule height 0.8pt}
		\end{tabular}
	\end{center}
	\caption{Kinematic self-similar solutions for the vacuum case. $\parallel$ and $\perp$ denote the parallel and the orthogonal cases, respectively. See text for the value of a parameter, $\alpha$.}
	\label{tablesol4}
\end{table}

\acknowledgements

We are very grateful to K.~Tomita, S.~Inutsuka and S.~Mineshige for useful discussions and J.~Overduin for his critical reading of our paper. We would also like to thank K.~Maeda for continuous encouragement. H.M. is grateful to the Yukawa Institute for Theoretical Physics at Kyoto University for hospitality received during this work.
This work was partly supported by the 
Grant-in-Aid for Scientific Research (Nos. 05540 and 11217)
from the Japanese Ministry of
Education, Culture, Sports, Science and Technology.

\appendix
\section{Possible equations of state}
\label{sec:possibleeos}

We have seen that the energy density and pressure 
are decomposed into one or two parts if the fluid 
is contained in a spherically symmetric spacetime 
which admit a kinematic self-similar vector. 
Here we consider what kind of equations of state
is possible if the energy density and pressure
are decomposed into the one or two parts.
Since the following analysis based only on the form of the energy density and pressure,
the equations of state may be restricted further 
if one demands that the matter content be a source 
of a spherically symmetric spacetime with the corresponding
kinematic self-similar vector. 

We consider a barotropic equation of state, i.e., 
\begin{equation}
p=f(\mu).
\label{eq:barotropic}
\end{equation}
It implies that the pressure is a function only of the energy density.
This class of equations of state is quite wide and 
it is often useful in many realistic and astrophysical situations.

\subsection{First kind}
For a tilted case, the energy density and pressure are expressed
in the following form 
\begin{eqnarray}
\mu&=&\frac{W(\xi)}{r^{2}}, \\
p&=&\frac{P(\xi)}{r^{2}},
\end{eqnarray}
where $\xi\equiv r/t$.
If we assume the barotropic equation of state, it is found that 
the function $f$ must be of the form
\begin{equation}
f(x)=kx,
\end{equation}
where $k$ is an arbitrary constant.  
This was proved by Cahill and Taub~\cite{CT1971}.

It can be easily proved that 
this result is also the case even for parallel and 
orthogonal cases.
\subsection{Second kind}
For a tilted case,  the 
energy density and pressure are decomposed into the 
following form
\begin{eqnarray}
\mu&=&\frac{W_{1}(\xi)}{r^{2}}+\frac{W_{2}(\xi)}{t^{2}}, \\
p &=& \frac{P_{1}(\xi)}{r^{2}}+\frac{P_{2}(\xi)}{t^{2}},  
\end{eqnarray}
where $\xi\equiv r/(\alpha t)^{1/\alpha}$ and $\alpha\ne 0, 1$.
The above can be rewritten as
\begin{eqnarray}
\mu&=&\frac{\tilde{W}_{1}(\xi)}{t^{2/\alpha}}+\frac{W_{2}(\xi)}{t^{2}}, \\
p &=& \frac{\tilde{P}_{1}(\xi)}{t^{2/\alpha}}+\frac{P_{2}(\xi)}{t^{2}},  
\end{eqnarray}
where $\tilde{W}_{1}\equiv \xi^{-2}\alpha^{-2/\alpha}W_{1}$
and $\tilde{P}_{1}\equiv \xi^{-2}\alpha^{-2/\alpha}P_{1}$.
Hereafter we will omit the tildes for simplicity.

Then, we can write down equation~(\ref{eq:barotropic}) as
\begin{equation}
\frac{P_{1}(\xi)}{t^{\alpha/2}}+
\frac{P_{2}(\xi)}{t^{2}}=
f\left(\frac{W_{1}(\xi)}{t^{\alpha/2}}
+\frac{W_{2}(\xi)}{t^{2}}\right).
\end{equation}
A partial differentiation with respect to $t$ leads to
\begin{equation} 
-\frac{2}{\alpha}\frac{P_{1}(\xi)}{t^{(2/\alpha) +1}}
-2\frac{P_{2}(\xi)}{t^{3}}
=f'\left(\frac{W_{1}(\xi)}{t^{\alpha/2}}
+\frac{W_{2}(\xi)}{t^{2}}\right)
\cdot
\left(-\frac{2}{\alpha}\frac{W_{1}(\xi)}{t^{(2/\alpha) +1}}
-2\frac{W_{2}(\xi)}{t^{3}}
\right),
\end{equation}
where and hereafter in this section a prime denotes
a derivative with respect to the argument.
A partial differentiation with respect to $\xi$ leads to
\begin{equation}
\frac{P_{1}'(\xi)}{t^{\alpha/2}}+
\frac{P_{2}'(\xi)}{t^{2}}=
f'\left(\frac{W_{1}(\xi)}{t^{\alpha/2}}
+\frac{W_{2}(\xi)}{t^{2}}\right)
\cdot
\left(\frac{W_{1}'(\xi)}{t^{\alpha/2}}
+\frac{W_{2}'(\xi)}{t^{2}}\right).
\end{equation}
From the above two equations, we can eliminate $f'$
as 
\begin{equation}
\left(-\frac{2}{\alpha}\frac{P_{1}(\xi)}{t^{(2/\alpha) +1}}
-2\frac{P_{2}(\xi)}{t^{3}}\right)
\left(\frac{W_{1}'(\xi)}{t^{\alpha/2}}
+\frac{W_{2}'(\xi)}{t^{2}}\right)
=\left(-\frac{2}{\alpha}\frac{W_{1}'(\xi)}{t^{(2/\alpha) +1}}
-2\frac{W_{2}'(\xi)}{t^{3}}
\right)
\left(\frac{P_{1}'(\xi)}{t^{\alpha/2}}+
\frac{P_{2}'(\xi)}{t^{2}}\right).
\end{equation}
Since $\alpha\ne 0,1$, the following relations must be satisfied.
\begin{eqnarray}
P_{1}W_{1}&=&W_{1}P_{1}', 
\label{eq:1}\\
P_{2}W_{2}&=&W_{2}P_{2}, 
\label{eq:2}\\
P_{1}W_{2}'+\alpha P_{2}W_{1}'&=& W_{1}P_{2}'+\alpha W_{2}P_{1}'.
\label{eq:3}
\end{eqnarray}

For $W_{1}W_{2}\ne 0$, we can integrate equations~(\ref{eq:1})
and (\ref{eq:2}) as
\begin{eqnarray}
P_{1}&=&k_{1}W_{1}, \\
P_{2}&=&k_{2}W_{2},
\end{eqnarray}
respectively, where $k_{1}$ and $k_{2}$ are constants of integration.
Substituting the above into equation~(\ref{eq:3}), we obtain
\begin{equation}
(k_{1}-k_{2})\left(-\frac{W_{2}'}{W_{2}}+\alpha \frac{W_{1}'}{W_{1}}\right)=0.
\end{equation}
If $k_{1}=k_{2}\equiv k$ is satisfied, we can find $f(x)=kx$.
If not, we obtain
\begin{equation}
W_{2}=C W_{1}^{\alpha},
\end{equation}
where $C$ is a constant of integration.
Then, we find
\begin{equation}
k_{1}x+k_{2}Cx^{\alpha}=f(x+Cx^{\alpha}).
\end{equation}
This equation of state is interpreted as
a mixture of two comoving fluids with linear equations 
of state $p_{1}=k_{1}\mu_{1}$ and $p_{2}=k_{2}\mu_{2}$
with the relation $\mu_{2}=C\mu_{1}^{\alpha}$, where
$p=p_{1}+p_{2}$ and $\mu=\mu_{1}+\mu_{2}$.
This equation of state is also interpreted as
a mixture of two comoving fluids with power-law equations
of state $p_{1}=k_{1}C^{-1/\alpha}\mu_{1}^{1/\alpha}$ and 
$p_{2}=k_{2}C^{\alpha}\mu_{2}^{\alpha}$ with the relation
$\mu_{1}=C\mu_{2}^{\alpha}$.

Next we consider the case in which $W_{1}$=0.
If $W_{2}\ne 0$, then we obtain $f(x)=kx+Cx^{1/\alpha}$,
where $k$ and $C$ are constants of integration.
If $W_{2}=0$, $P_{1}=P_{2}=0$ must be satisfied,
which results in a vacuum spacetime.

Finally we consider the case in which $W_{2}=0$.
If $W_{1}\ne 0$, then we obtain $f(x)=kx+Cx^{\alpha}$,
where $k$ and $C$ are constants of integration.

The above results are summarized by a class of functions
\begin{equation}
k_{1}x+k_{2}x^{\alpha}=f(C_{1}x+C_{2}x^{\alpha}),
\end{equation}
where $k_{1}$, $k_{2}$, $C_{1}$ and $C_{2}$
are arbitrary constants.
It is seen that 
the above class of equations of state 
includes EOS (I) for $k_{2}=C_{1}=0$ and $k_{1}C_{2}\ne 0$
or for $k_{1}=C_{2}=0$ and $k_{2}C_{1}\ne 0$,
EOS (II) for $k_{2}=0$ and $k_{1}C_{1}C_{2}\ne 0$
or for $k_{1}= 0$ and $k_{2}C_{1}C_{2}\ne 0$,
and EOS (III) 
for $k_{1}C_{2}=k_{2}C_{1}$, $(k_{1},k_{2})\ne (0,0)$ and 
$(C_{1},C_{2})\ne (0,0)$.

It can be easily proved that this result is also the case even
for parallel and orthogonal cases.
 
\subsection{Zeroth kind}
For a tilted case, the energy density and pressure are
decomposed into the following form
\begin{eqnarray}
\mu&=&\frac{W_{1}(\xi)}{r^{2}}+W_{2}(\xi), \\
p &=& \frac{P_{1}(\xi)}{r^{2}}+P_{2}(\xi),  
\end{eqnarray}
where $\xi=r/e^{t}$.
The barotropic equation of state
requires the following relations
\begin{eqnarray} 
W_{1}P_{1}'&=&P_{1}W_{1}', \\ 
W_{1}P_{2}'&=&P_{1}W_{2}'.
\end{eqnarray}

If $W_{1}\ne 0$, then we find
\begin{eqnarray}
P_{1}&=&kW_{1}, \\
P_{2}&=&kW_{2}+C,
\end{eqnarray}
where $k$ and $C$ are constants of integration.
Then, we obtain $f(x)=kx +C$.

If $W_{1}=0$, then $P_{1}=0$ or $W_{2}'=0$.
For $W_{2}'=0$, $W_{2}$ is constant, which implies $P_{1}=0$ and 
$P_{2}=\mbox{const}$. 
For $P_{1}=0$, we cannot specify the function $f$ any further.

The above results are summarized as follows:
if ${\cal L}_{\xi}p={\cal L}_{\xi}\mu=0$ is satisfied,
the equation of state cannot be specified any further 
from the two assumptions, while, if not,
the equation of state is given by
\begin{equation}
f(x)=kx +C,
\end{equation}
where $k$ and $C$ are arbitrary constants.

It can be easily proved that this result is also the case
even for parallel and orthogonal cases. 

\subsection{Infinite kind}
For a tilted case, the energy density and pressure are
decomposed into the following form
\begin{eqnarray}
\mu&=&\frac{W_{1}(\xi)}{t^{2}}+W_{2}(\xi), \\
p &=& \frac{P_{1}(\xi)}{t^{2}}+P_{2}(\xi),  
\end{eqnarray}
where $\xi=r/t$.

We can easily find that the result is 
completely the same as that for the zeroth kind.
If ${\cal L}_{\xi}p={\cal L}_{\xi}\mu=0$ is satisfied,
the equation of state cannot be specified any further 
from the two assumptions, while, if not,
the equation of state is given by
\begin{equation}
f(x)=kx +C,
\end{equation}
where $k$ and $C$ are arbitrary constants.

It can be easily proved that this result is also the case
even for an orthogonal case.
For a parallel case,
since ${\cal L}_{\xi}p={\cal L}_{\xi}\mu=0$ is satisfied, 
the equation of state cannot be specified any further.


\begin{thebibliography}{99}
\bibitem{carr1999}
  B.J.~Carr,
  in Proceedings of the Ninth Workshop on General Relativity and Gravitation, 
  Hiroshima, Japan, 1999, edited by Y. Eriguchi {\it et al}., p. 425, gr-qc/0003009. 

\bibitem{penston1969}
  M.V.~Penston,
  Mon. Not. R. Astr. Soc.
  {\bf 144} (1969), 425.

\bibitem{larson1969}
  R.B.~Larson,
  Mon. Not. R. Astr. Soc.
  {\bf 145} (1969), 271.

\bibitem{shu1977}
  F.H.~Shu,
  Astrophys. J.
  {\bf 214} (1977), 488.

\bibitem{hunter1977}
  C.~Hunter, 
  Astrophys. J.
  {\bf 218} (1977), 834.

\bibitem{ti1999}
  T.~Tsuribe and S.~Inutsuka,
  Astrophys. J.
  {\bf 526} (1999), 307.

\bibitem{hn1997}
  T.~Hanawa and K.~Nakayama,
  Astrophys. J.
  {\bf 484} (1997), 238.

\bibitem{hm2000a}
  T.~Hanawa and T.~Matsumoto,
  Publ. Astron. Soc. Jpn. 
  {\bf 52} (2000), 241. 

\bibitem{hm2000b}
  T.~Hanawa and T.~Matsumoto,
  Astrophys. J.
  {\bf 521} (2000), 703.

\bibitem{fc1993}
  P.N.~Foster and R.A.~Chevalier,
  Astrophys. J.
  {\bf 416} (1993), 303.

\bibitem{mh2001}
  H.~Maeda and T.~Harada,
  Phys. Rev. D
  {\bf 64} (2001), 124024. 

\bibitem{yahil1983}
  A.~Yahil,
  Astrophys. J.
  {\bf 265} (1983), 1047.

\bibitem{ss1988}
  Y.~Suto and J.~Silk,
  Astrophys. J.
  {\bf 326} (1988), 527.

\bibitem{CT1971}
  M.E.~Cahill and A.H.~Taub,
  Commun. Math. Phys.
  {\bf 21} (1971), 1.

\bibitem{op1987}
  A.~Ori and T.~Piran,
  Phys. Rev. Lett.
  {\bf 59} (1987), 2137.

\bibitem{op1988}
  A.~Ori and T.~Piran,
  Gen. Relat. Gravit.
  {\bf 20} (1988), 7.

\bibitem{op1990}
  A.~Ori and T.~Piran,
  Phys. Rev. D
  {\bf 42} (1990), 1068.

\bibitem{hm2001}
  T.~Harada and H.~Maeda,
  Phys. Rev. D
  {\bf 63} (2001), 084022.

\bibitem{harada2001}
  T.~Harada,
  Class. Quantum Grav.
  {\bf 18} (2001), 4549.

\bibitem{harada1998}
  T.~Harada,
  Phys. Rev. D
  {\bf 58} (1998), 104015.

\bibitem{hin2002}
  T.~Harada, H.~Iguchi and K.~Nakao,
  Prog. Theor. Phys.
  {\bf 107} (2002), 449.

\bibitem{ch}
  B.~Carter and R.N.~Henriksen,
  Ann. Phys. (Paris)
  {\bf 14} (1989), 47; 
  J. Math. Phys.
  {\bf 32} (1991), 2580.

\bibitem{tomita}
  K.~Tomita,
  Prog. Theor. Phys.
  {\bf 66} (1981), 2025;
  Suppl. Prog. Theor. Phys.
  {\bf 70} (1981), 286;
  Gen. Relat. Gravit.
  {\bf 29} (1997), 815.

\bibitem{carr2000}
  B.J.~Carr,
  Phys. Rev. D
  {\bf 62} (2000), 044022.

\bibitem{cc2000}
  B.J.~Carr and A.A.~Coley,
  Phys. Rev. D
  {\bf 62} (2000), 044023.

\bibitem{gnu1998}
  M.~Goliath, U.~Nilsson and C.~Uggla,
  Class. Quantum Grav.
  {\bf 15} (1998), 167;
  ibid.
  {\bf 15} (1998), 2841.

\bibitem{coley1991}
  A.A.~Coley,
  Class. Quantum. Grav.
  {\bf 8} (1991), 955.

\bibitem{mcintosh1975}
  C.B.G.~McIntosh,
  Gen. Relat. Gravit.
  {\bf 7} (1975), 199.

\bibitem{coley1997}
  A.A.~Coley,
  Class. Quantum Grav.
  {\bf 14} (1997), 87.

\bibitem{bc1998}
  P.M.~Benoit and A.A.~Coley,
  Class. Quantum Grav.
  {\bf 15} (1998), 2397.

\bibitem{sbc2001}
  A.M.~Sintes, P.M.~Benoit and A.A.~Coley,
  Gen. Relat. Gravit.
  {\bf 33} (2001), 1863.

\bibitem{blvw2002}
  C.F.C.~Brandt, L.-M.~Lin, J.F.~Villas da Rocha, and A.Z.~Wang,
  Int. J. Mod. Phys. D
  {\bf 11} (2002), 155.

\bibitem{mhio2002}
  H.~Maeda, T.~Harada, H.~Iguchi and N.~Okuyama,
  Phys. Rev. D
  {\bf 66} (2002), 027501.

\bibitem{st}
  S.L.~Shapiro and S.A.~Teukolsky,
  {\it Black Holes, White Dwarfs, and Newtron Stars}
  (Wiley, New York, 1983).

\bibitem{LL}
  L.D.~Landau and E.M.~Lifshitz,
  {\it The Classical Theory of Fields}
  (Pergamon, New York, 1975).

\bibitem{nariai}
   H.~Nariai,
   Sci. Rep. Tohoku Univ. Ser. 1 
   {\bf 35} (1951), 62.

\bibitem{mhio}
  H.~Maeda, T.~Harada, H.~Iguchi and N.~Okuyama,
  in preparation.

\bibitem{ph}
  J.~Ponce de Leon,
  J. Math. Phys.
  {\bf 29} (1988), 2479;
   Mon. Not. Roy. Astr. Soc.
  {\bf 250} (1991), 69;
  Gen. Relat. Gravit.
  {\bf 25} (1993), 865.

\bibitem{LTB}
   R.C.~Tolman,
   Proc. Natl. Acad. Sci. U.S.A. 
   {\bf 20} (1934), 169.\\
   H.~Bondi,
   Mon. Not. Roy. Astr. Soc. {\bf 107} (1947), 410.

\bibitem{coley1991}
  A.A.~Coley,
  Class. Quantum Grav.
  {\bf 8} (1991), 955.

\bibitem{mcintosh1975}
  C.B.G.~McIntosh,
  Gen. Relat. Gravit.
  {\bf 7} (1975), 199.

\end{thebibliography}
\end{document}